\documentclass[aps,prl,reprint,superscriptaddress]{revtex4-2}

\usepackage{graphicx}
\usepackage{epsfig}
\usepackage{amsmath,bbm,amssymb,bm,times}
\usepackage[colorlinks=true,linkcolor=blue,urlcolor=blue,citecolor=blue]{hyperref}

\usepackage{stmaryrd}
\usepackage[normalem]{ulem}
\usepackage{comment}
\usepackage[usenames,dvipsnames]{color}
  
\definecolor{dred}{rgb}{0.6, 0.0, 0.0}
\definecolor{dblue}{rgb}{0.0, 0.0, 0.6}

\definecolor{dgreen}{rgb}{0.0, 0.5, 0.0}

\newcommand{\abs}[1]{\left\lvert#1\right\rvert}

\begin{document}

\title{Bulk-Boundary Correspondence in Ergodic and Nonergodic One-Dimensional Stochastic Processes}

\author{Taro Sawada}
\email{sawada@noneq.t.u-tokyo.ac.jp}
\affiliation{Department of Applied Physics, The University of Tokyo, 7-3-1 Hongo, Bunkyo-ku, Tokyo 113-8656, Japan}
\author{Kazuki Sone}
\affiliation{Department of Applied Physics, The University of Tokyo, 7-3-1 Hongo, Bunkyo-ku, Tokyo 113-8656, Japan}
\affiliation{Department of Physics, University of Tsukuba, Tsukuba, Ibaraki 305-8571, Japan}
\author{Kazuki Yokomizo}
\affiliation{Department of Physics, The University of Tokyo, 7-3-1 Hongo, Bunkyo-ku, Tokyo 113-0033, Japan}
\author{Yuto Ashida}
\affiliation{Department of Physics, The University of Tokyo, 7-3-1 Hongo, Bunkyo-ku, Tokyo 113-0033, Japan}
\affiliation{Institute for Physics of Intelligence, The University of Tokyo, 7-3-1 Hongo, Tokyo 113-0033, Japan}
\author{Takahiro Sagawa}
\affiliation{Department of Applied Physics, The University of Tokyo, 7-3-1 Hongo, Bunkyo-ku, Tokyo 113-8656, Japan}
\affiliation{Quantum-Phase Electronics Center (QPEC), The University of Tokyo, 7-3-1 Hongo, Bunkyo-ku, Tokyo 113-8656, Japan}
\date{\today}
\begin{abstract}
    Bulk-boundary correspondence is a fundamental principle in topological physics.
    In recent years, there have been considerable efforts in extending the idea of geometry and topology to classical stochastic systems far from equilibrium.
    However, it has been unknown whether or not the bulk-boundary correspondence can be extended to the steady states of stochastic processes accompanied by additional constraints such as the conservation of probability.
    The present study reveals the general form of bulk-boundary correspondence in classical stochastic processes.
    Specifically, we prove a correspondence between the winding number and the number of localized steady states in both ergodic and nonergodic systems.
    Furthermore, we extend the argument of the bulk-boundary correspondence to a many-body stochastic model called the asymmetric simple exclusion process (ASEP).
    These results would provide a guiding principle for exploring topological origin of localization in various stochastic processes including biological systems.
\end{abstract}
\maketitle
\textit{Introduction.---}
As illuminated by the quantum Hall effect \cite{Klitzing1980, TKNN1982}, topological localization phenomena have been intensively explored in condensed matter physics.
Bulk-boundary correspondence is a key principle in topological materials stating that bulk topological invariants correspond to the number of the localized modes under the open boundary conditions (OBC). Those localization phenomena are robust against disorders since the topological invariants remain unchanged under continuous deformation of a Hamiltonian.
It is noteworthy that band topology has been further extended to non-Hermitian phenomena \cite{Kato1966, PhysRevB.97.121401, PhysRevLett.121.026808, ShenZhenFu2018, GongAshida2018}, such as the non-Hermitian skin effect \cite{Yao2018, YokomizoMurakami, NatPhys.16.747, PNAS.117.29561, NatPhys.16.761, Weidemann2020}.

On another front, recent studies have revealed the topological aspects of classical systems modeled by stochastic processes \cite{Sinitsyn2007, Jie2013, Murugan2017, PNAS.115.39, SciRep.11.888, Tang2021, Mahault2022, PhysRevE.104.025003, SciRep.12.560, Mehta2022, arXiv:2302.11503}.
We note that classical stochastic processes have significance in various fields such as stochastic thermodynamics, large deviation, and counting statistics \cite{vanKampen1992, SeifertReview2012, LebowitzSpohn2003, PhysRevB.67.085316}, and have attracted much attention in light of recent progress in experimental techniques \cite{PhysRevX.7.021051, NatPhys.6.988}.
A seminal work \cite{Murugan2017} indicated the correspondence between a topological index and localization of the steady state at the boundary between two different one-dimensional ergodic stochastic processes.
However, the general bulk-boundary correspondence for one-dimensional stochastic processes is still unestablished, especially given that nonergodic systems may have hidden symmetries and corresponding conserved quantities.

\begin{figure}[t]
        \centering
        \includegraphics[width=85mm, bb=0 0 620 180, clip]{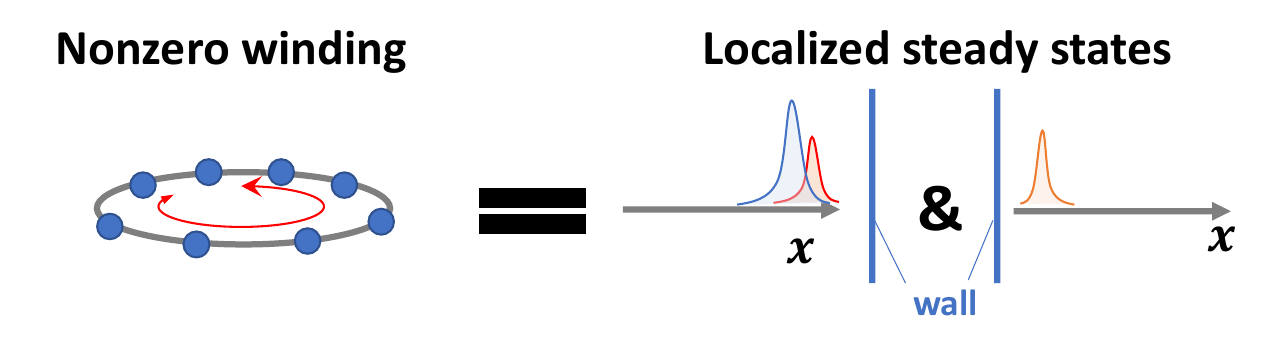}
        \caption{
        Schmatic of the bulk-boundary correspondence in stochastic processes.
        The left figure shows the nonzero winding number of a nonergodic system under the periodic boundary condition.
        The right figure shows the localized steady states under the left and the right semi-infinite boundary conditions (SIBCs).
        Our bulk-boundary correspondence states that a nonzero winding number must accompany with localized steady states under the SIBC or the open boundary condition.
        This schematic corresponds to the situation where $w = 1$, $N_L = 2$, and $N_R = 1$.
        }
        \label{fig: Schematic}
    \end{figure}

In this Letter, we reveal the general form of the bulk-boundary correspondence of one-dimensional stochastic processes including nonergodic cases with multiple steady states.
Specifically, we prove the bulk-boundary correspondence
    \begin{align} 
        w = N_L - N_R, \label{eqn: main statement}
    \end{align}
where $w$ is the bulk winding number and $N_L - N_R$ is difference of the numbers of left- and right-localized steady states as shown in Fig.~\ref{fig: Schematic}.
Since our result \eqref{eqn: main statement} can be applied to nonergodic processes, $w$ can take arbitrary integers, making sharp contrast to the ergodic case where $w$ only takes $0,\pm 1$ \cite{Murugan2017}.
In addition, we numerically show that the localized steady state has robustness against disorders, as a consequence of topological origin of localized steady states.

While we employ the non-Bloch band theory \cite{Yao2018, YokomizoMurakami} to prove Eq.~\eqref{eqn: main statement}, our result is unique to classical stochastic processes and does not fall into the general properties of non-Hermitian systems.
The key observation for the proof is that bulk current coincides with boundary current in non-Bloch waves.
Furthermore, we discuss the extension of the bulk-boundary correspondence to many-body stochastic systems, by considering the asymmetric simple exclusion process (ASEP) \cite{EPL.26.7,PhysRevLett.95.240601,JStatMech.2009.P07017}, describing zeolites \cite{PhysRevLett.76.2762}, traffic flows \cite{RevModPhys.73.1067}, protein synthesis \cite{RevModPhys.85.135, RevModPhys.91.045004}, and quantum-dot chain \cite{PhysRevB.81.045317}. 
Our results would provide the topological design principles of stochastic devices where the robustness suppresses the effect of the noise and disorder.

\textit{Setup.---}
We consider a general one-dimensional stochastic process described by a master equation $\frac{\mathrm{d}}{\mathrm{dt}}\lvert p(t) \rangle = W\lvert p(t) \rangle$
with the spatial system size $L$ and the internal degrees of freedom $K$. Here, $W$ is the transition-rate matrix, and $\lvert p(t) \rangle = \sum_{n=1}^L\sum_{\sigma=1}^K p(n,\sigma,t)\lvert n, \sigma\rangle$ is the probability distribution of the system, with $p(n,\sigma,t)$ being the probability that system is in the position $n$ and the internal degree of freedom $\sigma$.
We write the components of $W$ as $W_{nm;\sigma\nu} = \langle n,\sigma | W | m,\nu \rangle$.
We note that while under the PBC the system has complete translation invariance and no boundary, under the SIBC or the OBC the system lacks the complete translation invariance and has one or two-sided boundaries.
Moreover, the diagonal terms of the transition-rate matrix at the open boundary are different from those in the bulk due to the probability-preserving constraint.

In the following, we focus on spatially local transition-rate matrices and define the hopping range $l_0$ as the minimum integer so that $W_{nm;\sigma\nu}=0$ holds true for any $n,m,\sigma,\nu$ whenever $\abs{n-m}>l_0$.
Moreover, we assume the translation invariance $W_{nm;\sigma\nu} = W_{(n-m),0;\sigma\nu}$, where the value of $W_{(n-m),0;\sigma\nu}$ is determined by $(n-m), \sigma, \nu$.
In order to calculate the bulk eigenstates, we define $W^\lambda(k) := \sum_{n} W_{n0;\sigma\nu}e^{ikn}e^{\lambda n}$ and call it as the non-Bloch Hamiltonian of the transition-rate matrix.
By letting $\beta = e^{\lambda + ik}$, we write $W^\lambda(k)$ as $W(\beta)$.

In the following, we do not assume the ergodicity of stochastic processes.
We note that every ergodic stochastic process has a unique steady state with zero eigenvalue because of the Perron-Frobenius theorem \cite{Bhatia1997}, while nonergodic processes can have more than one steady states.

\textit{Winding numbers of stochastic processes.---}
To characterize the topology of stochastic processes, we use the winding number \cite{PhysRevLett.132.046602}.
Firstly, we consider $w_\pm := \lim_{\lambda \to \pm 0} w_\lambda$ with $w_\lambda:=(2\pi i)^{-1}\int_0^{2\pi} \frac{d}{dk}\log(\det(W^\lambda(k)))\mathrm{d}k$ and $W^\lambda(k)$ being the non-Bloch Hamiltonian.
Then, we define the winding number of a stochastic process as $w := w_+ + w_-$.
In an ergodic system, $\abs{w_\pm}$ is less than or equal to one since only one band contains the zero spectrum.
We note that this definition is analogous to the winding number used in the general theory of non-Hermitian topology \cite{GongAshida2018}.
However, we utilize the imaginary-gauge transformation to define the topological invariant around zero eigenvalue, which is undefined in the conventional non-Hermitian topology because a bulk spectrum of the transition-rate matrix contains the zero eigenvalue.
We have previously shown that in ergodic systems, the nonzero winding numbers always correspond to the nonzero derivative of the band that contains the zero eigenvalue \cite{PhysRevLett.132.046602}.

\textit{Bulk-boundary correspondence in stochastic processes.---}
The goal of this Letter is to show the bulk-boundary correspondence in stochastic processes, i.e., the correspondence between the winding number and the number of the localized steady states. Specifically, we prove the following theorem, whose sketch of the proof is provided later.

\textbf{Theorem:} \textit{Consider a translationally invariant one-dimensional stochastic process, which is not necessarily ergodic.
Let $w$ be its winding number and $N_L$ (resp. $N_R$) be the number of its zeromode under the left (resp. right) SIBC.
Then, $w$ is equal to their difference; $w=N_L-N_R$.}

For the asymmetric random walk $W(k) = a e^{-ik} + b e^{ik} -(a+b)$, this theorem indicates that the steady state is localized to the left (resp. right) boundary when $w = +1$ (resp. $w = -1$).
$w = +1$ and $w = -1$ correspond to the parameter regions $a<b$ and $a>b$, respectively.

Our results show the bulk-boundary correspondence unique to stochastic processes, while 
our theorem takes a similar form to conventional bulk-boundary correspondence in which bulk topological invariants correspond to the number of localized eigenstates.
In fact, while the general theory of non-Hermitian topological phase provides information about the number of localized modes at eigenvalues within the bulk bands, our theorem provides the number of localized modes at eigenvalues outside of the bulk bands.
Moreover, such localized modes only appear after imposing constraint of the probability preservation unique to stochastic processes.

We note that we can extend the above theorem to the case of OBC in which both sides of the system are open in a slightly weaker form (see SM for details).
The reason why the statements become weaker under the OBC is that the concepts of left- or right-localized steady states are unclear when the system has boundaries on both sides.

\begin{figure}[t]
        \centering
        \includegraphics[width=85mm, bb=0 0 450 230, clip]{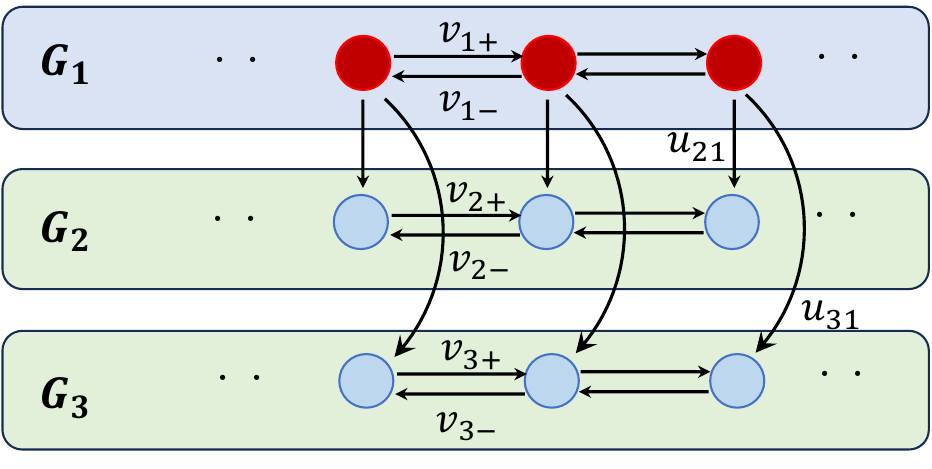}
        \caption{
        Transition diagram of the nonergodic model \eqref{eqn: nonergodic model} used in the numerical calculation, which has two steady states.
        Each $G_i$ ($i=1,2,3$) represents the strongly connected components of the model. 
        $G_2$ and $G_3$ are the sink components which is defined in the proof of nonergodic case.
        }
        \label{fig: nonergodic example transition diagram}
    \end{figure}

\textit{Example and robustness.---}
We numerically demonstrate our bulk-boundary correspondence in a nonergodic model (Fig.~\ref{fig: nonergodic example transition diagram}): 
    \begin{align}
        &\frac{\mathrm{d}}{\mathrm{d}t} \bm{p}(n,t) \nonumber
        = \begin{pmatrix}
            v_{1+} & 0 & 0 \\
            0 & v_{2+} & 0 \\
            0 & 0& v_{3+}
        \end{pmatrix}\bm{p}(n-1,t) \nonumber\\
        &+ \begin{pmatrix}
            v_{1-} & 0 & 0 \\
            0 & v_{2-} & 0 \\
            0 & 0& v_{3-}
        \end{pmatrix}\bm{p}(n+1,t) 
         + \begin{pmatrix}
            d_1 & 0 & 0 \\
            u_{21} & d_2 & 0 \\
            u_{31} & 0 & d_3
        \end{pmatrix}\bm{p}(n,t), \label{eqn: nonergodic model}
    \end{align}
where $\bm{p}(n,t) = (p(n,1,t), p(n,2,t), p(n,3,t))^\top$ is the probability vector, $d_1 = -(u_{21} + u_{31} + v_{1+} + v_{1-})$, $d_2 = -(v_{2+} + v_{2-})$, $d_3 = -(v_{3+} + v_{3-})$ are diagonal losses originated from the probability-preserving constraint.
The nonergodicity is confirmed since the restriction $p(n,1,t) = 0$ gives the diagonal transition-rate matrix.

In particular, we verify that the localization of the steady states is robust against disorders.
To confirm the robustness, we impose the off-diagonal disorder $W_{nm;\sigma\nu} \mapsto \tilde{W}_{nm;\sigma\nu} + \Delta_{nm;\sigma\nu}$ ($(n,\sigma)\neq (m,\nu)$) where each $\Delta_{nm;\sigma\nu}$ is randomly generated from the uniform distribution on $[-\delta, \delta]$ ($\delta > 0$).
We calculate the mean and the variance of the probability distribution of the steady states from $1000$ samples.

\begin{figure}[t]
        \centering
        \includegraphics[width=85mm, bb=0 0 800 810, clip]{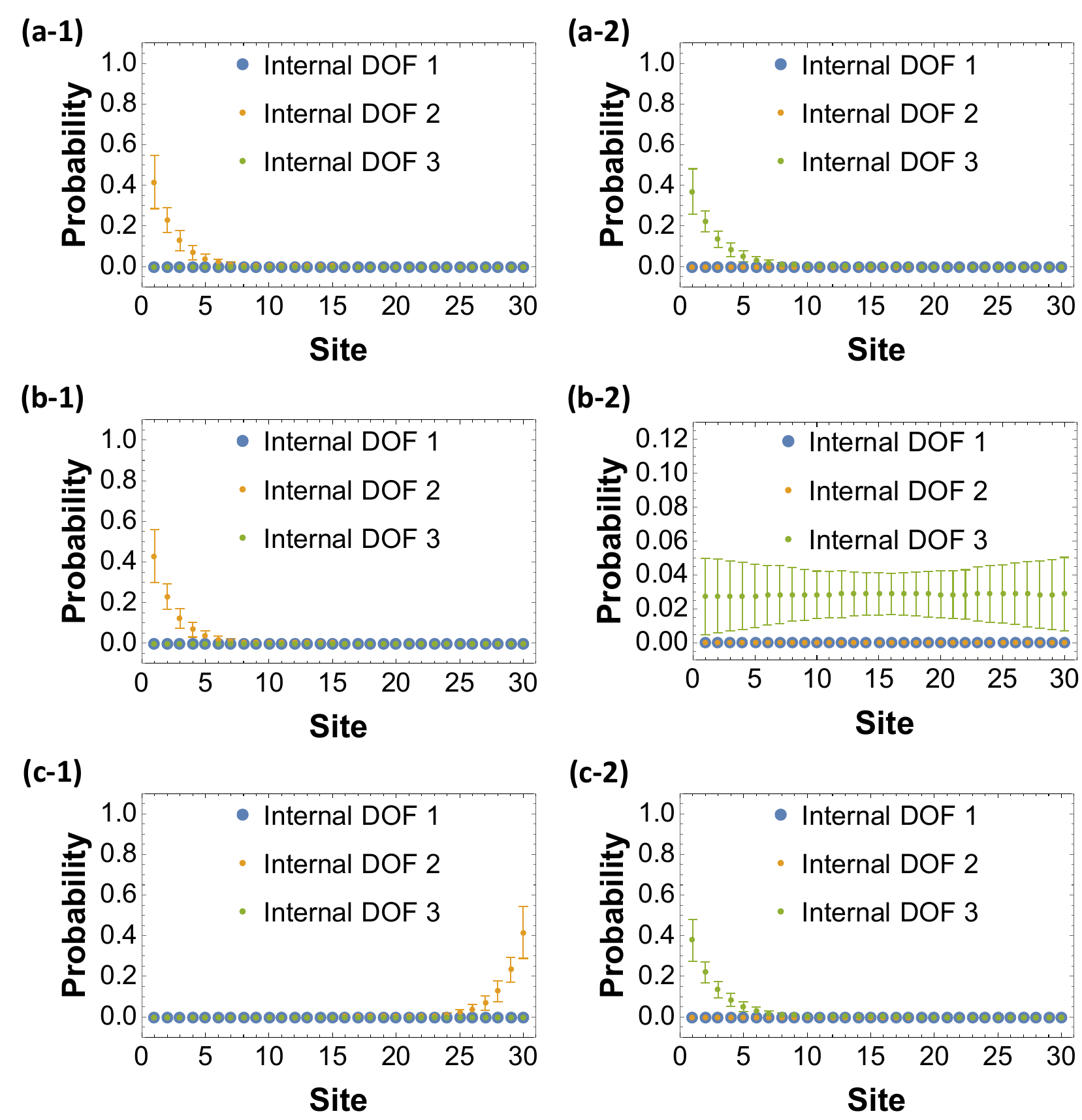}
        \caption{
        Numerical calculation of steady states and their robustness against disorder in the nonergodic stochastic process under the OBC.
        The DOF in legends is abbreviation for degrees of freedom.
        We set the system size as $L=30$ and the disorder amplitude as $\delta = 0.25$.
        Subnumbers (1,2) represent linearly independent steady states obtained within the same parameter set.
        The dots (error bars) represent the mean values (variance) of the probability amplitude with off-diagonal disorders.
        The localization direction of the steady states correspond to the value of the winding number (a) $w=2$, (b) $w=1$, and (c) $w=0$.
        Parameters used are ($u_{21},u_{31},v_{1+},v_{1-},v_{2+},v_{2-},v_{3+},v_{3-}$) = (a) (1, 2, 0.7, 0.6, 0.4, 0.7, 0.5, 0.8), (b) (1, 2, 0.7, 0.6, 0.7, 0.4, 0.8, 0.8), (c) (1, 2, 0.7, 0.6, 0.7, 0.4, 0.5, 0.8).
        }
        \label{fig: nonergodic example steady states}
    \end{figure}

Figure~\ref{fig: nonergodic example steady states} shows the results of the numerical calculation for the nonergodic model \eqref{eqn: nonergodic model}.
In accordance with the theorem, the system has the localized steady states corresponding to the winding number.
While Fig.~\ref{fig: nonergodic example steady states}(a) (resp. (b)) exhibits two (one) left-localized steady states corresponding to the winding number $w = 2$ (resp. $w = 1$), Fig.~\ref{fig: nonergodic example steady states}(c) exhibits one left- and one right-localized steady states corresponding to the relation $w = 0 = 1 - 1$.
The delocalized mode shown in Fig.~\ref{fig: nonergodic example steady states}(b-2) is highly affected by the disorder because of the Anderson localizaion \cite{PhysRev.109.1492}.
We note that the amplitude of the steady states are concentrated on a single internal degrees of freedom, denoted as $G_2$ or $G_3$.
This concentration of the steady state on a strongly connected component is proved later as Lemma 2.

\textit{Non-Bloch wave expansion of steady states.---}
To prove Theorem, we utilize the non-Bloch wave expansion of the steady state $|p_\textrm{ss}\rangle$ to detect the localization under the SIBC or the OBC:
    \begin{align}
    \langle n|p_\textrm{ss}\rangle &= \sum_j c_j (\beta_j)^n |\phi_j(\beta_j)\rangle, \label{eqn: non-Bloch wave expansion main text} \\
    W(\beta_j)\lvert\phi_j(\beta_j)\rangle &= 0, \label{eqn: bulk eigenequation}
    \end{align}
where $\lvert \phi_j(\beta_j) \rangle$ represents zero eigenvectors of the non-Bloch Hamiltonian $W(\beta)_{\sigma \nu} := \sum_{n} W_{n0;\sigma\nu} \beta^{-n}$.
$\abs{\beta_j} < 1$ (resp. $\abs{\beta_j} > 1$) indicates left-localized (resp. right-localized) waves, and $\abs{\beta_j} = 1$ indicates delocalized waves.
The non-Bloch expansion has been previously used in the non-Bloch band theory \cite{Yao2018, YokomizoMurakami} and we can justify this method by considering a transfer matrix \cite{Elaydi}.

Let $N^\lambda$ represent the number of solutions to $\det(W(\beta))=0$ under the constraint $|\beta| < e^\lambda$.
Then, we derive $w_\lambda = N^\lambda - l_0 K$ using the residue theorem.
We also obtain the inequality $w_- \leq w_+$ from this relation.
Moreover, $\beta = 1 $ is a trivial solution of $\det(W(\beta))=0$ since the equation $\sum_{\sigma} W(\beta = 1)_{\sigma \nu} = \sum_{n, \sigma}W_{n0;\sigma \nu} = 0$ holds true from the probability conservation, and it tells us that $W(\beta = 1)$ is also a transition-rate matrix.
Therefore, we obtain the stronger inequality $w_- < w_+$ since $\beta =1$ is counted in $N^+ = \lim_{\lambda\to +0} N^\lambda$ but not in $N^- = \lim_{\lambda\to -0} N^\lambda$.
We note that in ergodic systems with nonzero winding number, the delocalized component of the non-Bloch wave expansion (Eq.~\eqref{eqn: non-Bloch wave expansion main text}) is only the uniform wave $\beta = 1$ since the possible patterns of $(w_+, w_-)$ are only two: $(w_+, w_-)=(1,0),(0,-1)$, which means the unimodular solution of $\det(W(\beta))=0$ is unique.

\begin{figure}[t]
        \centering
        \includegraphics[width=85mm, bb=0 0 920 390, clip]{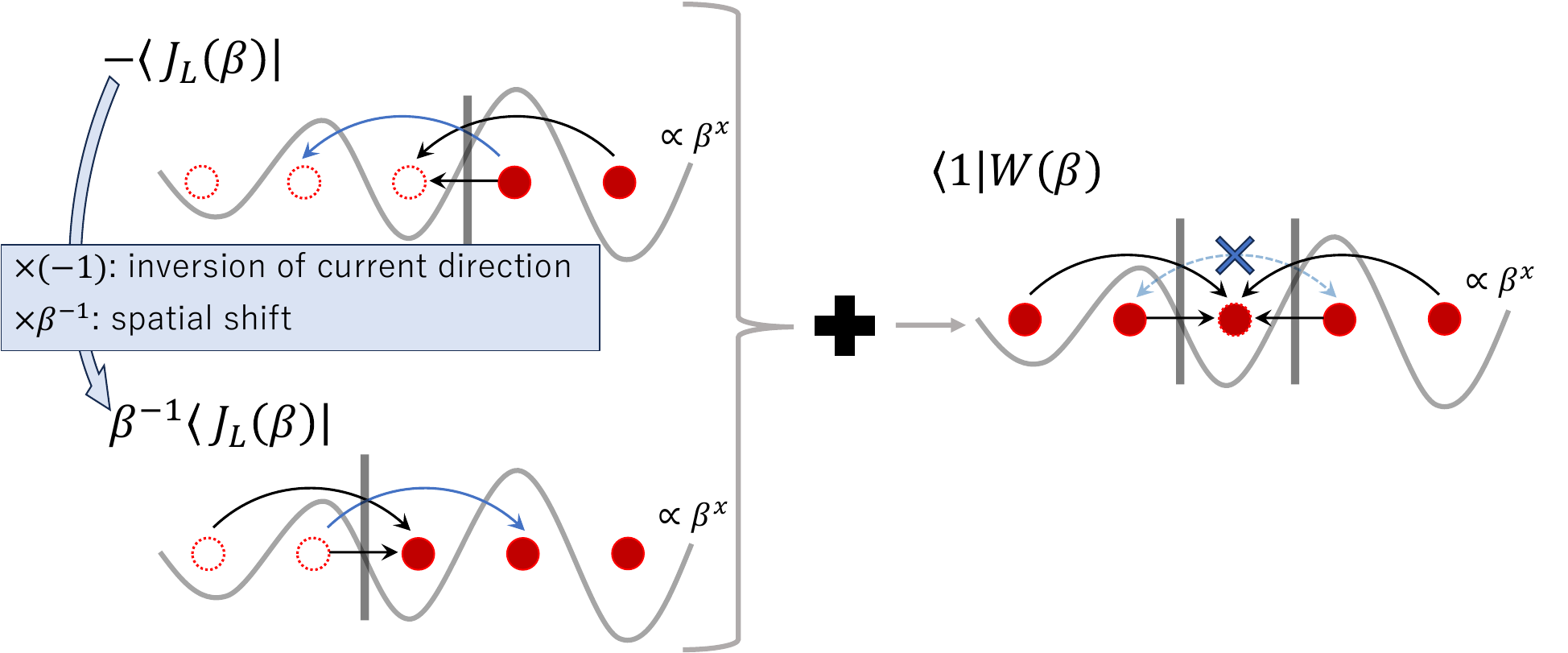}
        \caption{
        Graphical proof of the key observation.
        We prove the equality between bulk and boundary currents for the non-Bloch wave.
        The upper (resp. lower) left figure represent the inverted boundary current (resp. spatially shifted one).
        The right figure represents bulk current.
        One can confirm that the summation of left diagrams is equal to the right diagram and the same discussion works in general other models.
        }
        \label{fig: schematic bulk-boundary correspondence}
    \end{figure}

\textit{Outline of proof: ergodic case.---}
We give the outline of proof in the ergodic case.
Firstly, under the SIBC, we prove the correspondence between bulk currents and boundary currents, which we call the current bulk-boundary correspondence,
  \begin{equation}
    \langle 1 \rvert W(\beta) = (\beta^{-1} - 1)\langle J_L(\beta)\rvert, \label{eqn: correspondence btwn bulk and boundary currents}
  \end{equation}
regardless of the topology of the system, where the $\langle J_L(\beta)\rvert$ is the boundary current defined as
  \begin{align}
    &\langle J_L(\beta_j)\rvert_\nu \notag \\
    &= \sum_\sigma \sum_{m=1}^{l_0} \sum_{n=1}^{m} \left(W_{m,0;\sigma \nu}\left(\beta_j\right)^{-m} - W_{-m,0;\sigma \nu} \right)\left(\beta_j\right)^n. \label{eqn: definition of boundary current}
  \end{align}
$\langle J_L(\beta)\rvert$ is defined from the sum of the system of equations at the left boundary $\sum_{m=1}^{n+l_0} \sum_\nu \left(W_{nm;\sigma \nu} p_{\textrm{ss}}(m, \nu)- W_{mn;\sigma \nu}p_{\textrm{ss}} (n, \nu)\right) = 0$ ($n= 1, \ldots, l_0$, $\sigma = 1,\ldots, K$) which reads 
    \begin{equation}
        \sum_j c_j \langle J_L(\beta_j) | \phi_j(\beta_j) \rangle = 0. \label{eqn: boundary constraint}
    \end{equation}
The schematic of the Eq.~\eqref{eqn: correspondence btwn bulk and boundary currents} is given in Fig.~\ref{fig: schematic bulk-boundary correspondence}.
Since the spatial shift is equivalent to the multiplication by $\beta$ in a non-Bloch wave $p(n) \propto \beta^n$, Eq.~\eqref{eqn: correspondence btwn bulk and boundary currents} is the equality between the sum of the spatial currents in bulk and boundary.
We note that this current bulk-boundary correspondence is unique to stochastic processes since the balance of currents at the boundary originates from the constraints of probability conservation $\langle 1 \rvert W = 0$.

By the current bulk-boundary correspondence, we obtain $\langle J_L(\beta_j) | \phi_j(\beta_j) \rangle = (\beta_j^{-1} - 1)^{-1}\langle 1 | W(\beta_j) | \phi_j(\beta_j) \rangle = 0$ when $\beta_j \neq 1$, and $\langle J_L(\beta_j) | \phi_j(\beta_j) \rangle = (\partial_\beta E)_{\beta=1}$ when $\beta_j =1$.
Therefore, the boundary constraint \eqref{eqn: boundary constraint} reads $c_{j_0} (\partial_\beta E)_{\beta=1} = 0$ where $j_0$ is the index corresponding to $\beta_{j_0} = 1$.
Furthermore, since $(\partial_\beta E)_{\beta=1} \neq 0$ holds true in nonzero-winding-number systems, we obtain $c_{j_0}=0$, meaning that the zeromode is the composite of left- or right-localized waves.
Finally, we show that the steady states can be described by superpositions of left- or right-localized non-Bloch wave corresponding to the winding number.

\textit{Outline of proof: nonergodic case.---}
To prove a nonergodic case, we use several lemmas.
In those lemmas, we introduce graph-theoretic concepts related to directed graphs.
A directed graph induced by a transition-rate matrix $W$ is a directed graph whose connection matrix $G$ satisfies $G_{ij} = 0$ when $W_{ij} = 0$ or $i=j$ and $G_{ij} = 1$ otherwise.
Directed graphs are always decomposed into the strongly connected components, which physically corresponds to ergodic subsystems.
Decomposition of strongly connected components prohibit bidirectional hopping between any two strongly connected components $G_i$ and $G_j$ but allows unidirectional hopping from a site $s_i$ in $G_i$ to a site $s_j$ in $G_j$.
A strongly connected component is called sink component if it does not have such hopping directed to a site in another component.
Every sink component is an isolated subsystem, and therefore the transition-rate matrix restricted to a sink component also satisfies the probability-preserving constraint.

We prove the Lemma 1 on the structure of nonergodic systems.
It ensures that the subsystems are ergodic translationally invariant systems, where we have already shown the bulk-boundary correspondence.

\textbf{Lemma 1: } \textit{Let $W$ be the transition-rate matrix of a translationally invariant stochastic process, which is not necessarily ergodic.
Then, the number of zeromodes of $W$ increases in $O(L)$ with $L$ being the system size, or the sink components of the directed graph induced by $W$ also have translation invariance.}

Moreover, it is also possible to consider the winding number and the steady state in each sink component separately as summarized in the Lemma 2 and 3.
Therefore, the statement of the theorem also holds true in nonergodic cases.

\textbf{Lemma 2: } \textit{For every translationally invariant stochastic process $W$ which is not necessarily ergodic, the winding number $w$ is the sum of winding numbers $w_i$ calculated separately from sink components.}

\textbf{Lemma 3: } \textit{For every translationally invariant stochastic process $W$ which is not necessarily ergodic, the steady states have nonzero amplitude only on sites in the sink components.}

\textit{Extension to many-body case.---}
We next discuss that our bulk-boundary correspondence can be extended to many-body systems.
Specifically, we show that our bulk-boundary correspondence holds true in a simplest many-body stochastic process called the ASEP.
To formulate the bulk-boundary correspondence, we need to fix the particle number, which is conserved under the PBC and some OBC.
Then, we check the localization of the steady state under the OBC using the explicit formula obtained by the coordinate Bethe ansatz \cite{JStatMech.2009.P07017}.

Firstly, we define the winding number of the ASEP with fixed particle number $N$ in a similar manner to the one-particle case, replacing the winding number with the many-body winding number \cite{PhysRevB.105.165137}.
Since the ASEP under PBC with a fixed particle number is ergodic, the winding number takes the value $0$ or $\pm 1$.
Next, we observe the localization of the steady state in the ASEP under the OBC using the Bethe ansatz. Precisely, we check $P(x_1, \ldots, x_N) \propto (b/a)^{\sum_i x_i}$ where $P(x_1, \ldots, x_N)$ is the probability that each particle labeled by $i$ takes its position $x_i$.
Confirming that the winding number becomes nonzero in the case of $b/a\neq 1$, we conclude that the winding number corresponds to the signed number of the localized steady state under particle-number conservation.
We note that even under the OBC with boundary injection and ejection, the steady state is solved by the matrix Bethe ansatz \cite{JPhysMathGen.26.1493} and localization can be observed in some parameter region.
This fact suggests that it is possible to extend the definition of the winding numbers and our bulk-boundary correspondence to non-conservative many-body systems.

\textit{Discussion.---}
We showed the bulk-boundary correspondence between the winding number and the localized steady state in general translationally invariant one-dimensional stochastic processes.
The key observation is the correspondence between bulk and boundary currents for non-Bloch waves. 
We also numerically confirmed the robustness of the localization property of the steady states.
While our previous paper \cite{PhysRevLett.132.046602} considered the non-Hermitian skin effect in bulk modes, we here proved the localization of steady states, which are different from bulk modes.

One can experimentally realize one-particle systems and the localization phenomena using active matter \cite{Rupprecht2016, Klumpp2016, NatRevMolCellBiol.11.633, Alessandro2021, Yamauchi2020, Gangwal2008, NatCommun.12.4691} and cell adhesions \cite{Alessandro2021, Yamauchi2020}.
The localization in the ASEP can also be confirmed in traffic flows \cite{RevModPhys.73.1067}, protein synthesis \cite{RevModPhys.85.135, RevModPhys.91.045004}.
Moreover, the non-conservative ASEP with boundary injection and ejection can be designed by the quantum-dot chain under spatial voltage \cite{PhysRevB.81.045317}.

The topological localization proved in this research differs from the previously reported ones in the general theory of the non-Hermitian topology \cite{PhysRevLett.124.056802,Okuma2020,PhysRevLett.125.126402}.
Furthermore, under the OBC, the zero spectrum is outside of the bulk continuum spectrum determined by the GBZ condition $\abs{\beta_{l_0K}} = \abs{\beta_{l_0K+1}}$ \cite{Yao2018, YokomizoMurakami}.
Moreover, the localized steady states do not originate from the line-gap topology of the bulk.
For these reasons, the localized steady states given in this Letter are unique to stochastic processes.
We note that the delocalization is not topologically protected since the Anderson localizaion \cite{PhysRev.109.1492} occurs under the existence of infinitesimal disorder.

It is also noteworthy that our theoretical analysis broadens the utility of the non-Bloch band theory \cite{Yao2018, YokomizoMurakami} to the spectrum outside the bulk and shows that it can also capture localized modes unique to classical stochastic systems, which should be good platforms to investigate exotic non-Hermitian phenomena.
In addition, the extension of our bulk-boundary correspondence to higher-dimensional cases remains an important issue and may be conducted by the help of recent progresses \cite{PhysRevB.107.195112, NatCommun.13.2496, PhysRevX.14.021011} in the theory of general non-Hermitian systems.

\begin{acknowledgements}
\textit{Acknowledgements.---}
We thank Hosho Katsura, Daiki Nishiguchi and Zongping Gong for valuable discussions.
T. Sawada and K.S. are supported by World-leading Innovative Graduate Study Program for Materials Research, Information, and Technology (MERIT-WINGS) of the University of Tokyo.
K.S. is also supported by JSPS KAKENHI Grant Number JP21J20199.
K.Y. is supported by JSPS KAKENHI through Grant No. JP21J01409.
Y.A. acknowledges support from the Japan Society for the Promotion of Science through Grant No. JP19K23424 and from JST FOREST Program (Grant Number JPMJFR222U, Japan) and JST CREST (Grant Number JPMJCR23I2, Japan).
T. Sagawa is supported by JSPS KAKENHI Grant Number JP19H05796, JST CREST Grant Number JPMJCR20C1, JST ERATO-FS Grant Number JPMJER2204, and JST ERATO Grant Number JPMJER2302, Japan.
T. Sagawa is also supported by Institute of AI and Beyond of the University of Tokyo.
\end{acknowledgements}
\bibliography{Zeromode1DSP_TopoLocalization}

\widetext
\pagebreak
\begin{center}
\textbf{\large Supplementary Material for ``Bulk-Boundary Correspondence in Ergodic and Nonergodic One-Dimensional Stochastic Processes"}
\end{center}

\renewcommand{\theequation}{S\arabic{equation}}
\renewcommand{\thefigure}{S\arabic{figure}}
\renewcommand{\bibnumfmt}[1]{[S#1]}
\setcounter{equation}{0}
\setcounter{figure}{0}


\newcommand{\bra}[1]{\langle #1 \rvert}
\newcommand{\ket}[1]{\lvert #1 \rangle}
\newcommand{\braket}[2]{\langle #1 | #2 \rangle}

\subsection{Derivation of the boundary currents}
In this section, we derive the formula of the boundary current $\bra{J_L(\beta)}$ which acts on each non-Bloch wave in the expansion of the steady state
  \begin{align}
    \braket{n}{p_\textrm{ss}} &= \sum_j c_j \left(\beta_j\right)^n \ket{\phi(\beta_j)}, \\
    \textrm{i.e.} \braket{n, \tau = \sigma}{p_\textrm{ss}} &= \sum_j c_j \left(\beta_j\right)^n \phi(\beta_j)_\sigma.
  \end{align}
For the sake of notational simplicity, we write $\braket{x=n, \tau = \sigma}{p_\textrm{ss}}$ as $p_{\textrm{ss}}(n, \sigma)$ in the rest of this subsection.

\subsubsection{Comments on boundary conditions}
Before the calculation, we comment on the definitions of the semi-infinite boundary condition (SIBC) and the open boundary condition (OBC) modified to satisfy the constraint of the probability conservation.
When the system has boundaries, the diagonal loss terms at the boundary are modified through the relation $W_{nn;\sigma \sigma} = -\sum_{m\neq n}\sum_{\nu\neq\mu}W_{nm;\sigma \nu}$ since the hoppings directed towards outside of the boundary disappear.
We note that this kind of boundary conditions are called reflective (or reflecting) boundary condition in the references of the many-body physics \cite{EPL.26.7}.

We provide the example of the asymmetric random walk $H(k) = ae^{-ik} + be^{ik}$.
The master equation at the left (resp. right) boundary is $ \frac{\textrm{d}}{\textrm{d}t} p(1) = b p(2) - a p(1)$ (resp. $ \frac{\textrm{d}}{\textrm{d}t} p(L) = a p(L-1) - b p(L)$).
These equations result in diagonal losses $-a$ (resp. $-b$) at the left (resp. right) boundary.

\subsubsection{The bulk equation}
We note that each $\ket{\phi(\beta_j)}$ satisfies $W(\beta)\ket{\phi(\beta_j)}=0$, i.e., $\sum_{m=-l_0}^{l_0} \left(W_{0m;\sigma \nu} \beta_j^{m} - W_{m0;\sigma \nu}\right)\phi(\beta_j)_\nu = 0 $.
Therefore, by multiplying $c_j(\beta_j)^n$ and summing up with respect to $j$, we obtain 
  \begin{align}
    \sum_{m=n-l_0}^{n+l_0} \left(W_{nm;\sigma \nu} p_{\textrm{ss}}(m, \nu) - W_{mn;\sigma \nu}p_{\textrm{ss}}(n, \nu)\right) = 0 \label{eqn: bulk equation}
  \end{align}
for all $n$, $\sigma$.
We call it the bulk equation in the rest of this subsection.

\subsubsection{Calculation of the equations at the boundary}
Next, we explicitly write down the equations at the left boundary $\sum_{m=1}^{n+l_0} \sum_\nu \left(W_{nm;\sigma \nu} p_{\textrm{ss}}(m, \nu)- W_{mn;\sigma \nu}p_{\textrm{ss}}(n, \nu)\right) = 0$ $(n= 1, \ldots, l_0, \forall \sigma)$.
Taking the difference of it from the bulk equation \eqref{eqn: bulk equation}, we obtain
  \begin{align}
    \sum_{m=n-l_0}^{0} \sum_\nu \left(W_{nm;\sigma \nu} p_{\textrm{ss}}(m, \nu) - W_{mn;\sigma \nu}p_{\textrm{ss}}(n, \nu)\right) &= 0 \:(n= 1, \ldots, l_0, \forall \sigma).
  \end{align}
Rewriting it in terms of $\beta_j$, it reads
  \begin{align}
    \sum_j c_j \sum_\nu \sum_{m=n-l_0}^{0} \left( W_{n-m,0;\sigma \nu}\left(\beta_j\right)^m - W_{m-n,0;\sigma \nu}\left(\beta_j\right)^n \right) \phi(\beta_j)_\nu &= 0 \:(n= 1, \ldots, l_0, \forall \sigma).
  \end{align}
The $m$-summation part is transformed by the change of variable from $m$ to $\tilde{m}=n-m$ as
  \begin{align}
    \sum_{\tilde{m}=n}^{l_0} \left( W_{\tilde{m},0;\sigma \nu}\left(\beta_j\right)^{-\tilde{m}+n} - W_{-\tilde{m},0;\sigma \nu}\left(\beta_j\right)^n \right) \phi(\beta_j)_\nu &= 0 \:(n= 1, \ldots, l_0, \forall \sigma),
    \end{align}
which is equivalent to
    \begin{align}
    \sum_{\tilde{m}=n}^{l_0}  \left( W_{\tilde{m},0;\sigma \nu}\left(\beta_j\right)^{-\tilde{m}} - W_{-\tilde{m},0;\sigma \nu} \right)\left(\beta_j\right)^n \phi(\beta_j)_\nu &= 0 \:(n= 1, \ldots, l_0, \forall \sigma).
  \end{align}
Finally, we reach
  \begin{align}
    \sum_j c_j \sum_\nu \sum_{m=n}^{l_0} \left( W_{m,0;\sigma \nu}\left(\beta_j\right)^{-m} - W_{-m,0;\sigma \nu} \right)\left(\beta_j\right)^n \phi(\beta_j)_\nu &= 0 \:(n= 1, \ldots, l_0, \forall \sigma) \label{eqn: left SIBC}.
  \end{align}
Conducting the summation for all $n$ and $\sigma$ gives the desired formula
  \begin{align}
    \sum_j c_j \braket{J_L(\beta_j)}{\phi(\beta_j)} &= 0, \label{eqn: left boundary equation}\\
    \bra{J_L(\beta_j)}_\nu &= \sum_\sigma \sum_{n=1}^{l_0} \sum_{m=n}^{l_0} \left( W_{m,0;\sigma \nu}\left(\beta_j\right)^{-m} - W_{-m,0;\sigma \nu} \right)\left(\beta_j\right)^n.
  \end{align}
Furthermore, by changing the range of the summation
  \begin{align}
    \sum_{n=1}^{l_0} \sum_{m=n}^{l_0} = \sum_{m=1}^{l_0} \sum_{n=1}^{m},
  \end{align}
we obtain
  \begin{align}
    \bra{J_L(\beta_j)}_\nu &= \sum_\sigma \sum_{m=1}^{l_0} \sum_{n=1}^{m} \left( W_{m,0;\sigma \nu}\left(\beta_j\right)^{-m} - W_{-m,0;\sigma \nu} \right)\left(\beta_j\right)^n.
  \end{align}
which is nothing but Eq.~(6) in the main text.


\subsection{Equivalence of the left and the right boundary}
We also establish the equivalence $\bra{J_R(\beta)} = -\beta^L\bra{J_L(\beta)}$ of the left boundary current $\bra{J_L(\beta)}$ and the right boundary current $\bra{J_R(\beta)}$, the latter defined similarly to $\bra{J_L(\beta)}$ from the right boundary equation:
  \begin{align}
    \sum_j c_j \braket{J_R(\beta_j)}{\phi(\beta_j)} &= 0, \\
    \bra{J_R(\beta_j)}_\nu &= \sum_\sigma \sum_{n=1}^{l_0} \sum_{m=n}^{l_0} \left( W_{-m,0;\sigma \nu}\left(\beta_j\right)^{m} - W_{m,0;\sigma \nu} \right)\left(\beta_j\right)^{(L+1) - n} \\
    &= \sum_\sigma \sum_{m=1}^{l_0} \sum_{n=1}^{m} \left( W_{-m,0;\sigma \nu}\left(\beta_j\right)^{m} - W_{m,0;\sigma \nu} \right)\left(\beta_j\right)^{(L+1) - n}.
  \end{align}
The proof of equivalence is carried out by direct calculation:
  \begin{align}
    \bra{J_R(\beta_j)}_\nu &= \sum_\sigma \sum_{m=1}^{l_0} \sum_{n=1}^{m} \left( W_{-m,0;\sigma \nu}\left(\beta_j\right)^{m} - W_{m,0;\sigma \nu} \right)\left(\beta_j\right)^{(L+1) - n} \\
    &= \left(\beta_j\right)^{L}\sum_\sigma \sum_{m=1}^{l_0} \left(W_{-m,0;\sigma \nu} - W_{m,0;\sigma \nu}\left(\beta_j\right)^{-m} \right) \left(\sum_{n=1}^{m}\left(\beta_j\right)^{m+1 - n} \right) \\
    &= -\left(\beta_j\right)^{L}\sum_\sigma \sum_{m=1}^{l_0} \left( W_{m,0;\sigma \nu}\left(\beta_j\right)^{-m} - W_{-m,0;\sigma \nu} \right)\left( \sum_{n=1}^{m}\left(\beta_j\right)^{n} \right) \\
    &= -\left(\beta_j\right)^{L}\sum_\sigma \sum_{m=1}^{l_0} \sum_{n=1}^{m}\left( W_{m,0;\sigma \nu}\left(\beta_j\right)^{-m} - W_{-m,0;\sigma \nu} \right) \left(\beta_j\right)^{n}  \\
    &= -\left(\beta_j\right)^{L}\bra{J_L(\beta_j)}_\nu.
  \end{align}
Thanks to this equivalence, the current bulk-boundary correspondence also holds true for the right boundary as
    \begin{equation}
        \bra{1}W(\beta) = \beta^{-L}(1-\beta^{-1})\bra{J_R(\beta)}.
    \end{equation}
The following argument under the left SIBC also works under the right SIBC.

\subsection{The details in the proof of ergodic case}
We give the details in the proof of the theorem in the main text.

\textbf{Theorem:} \textit{Consider a translationally invariant one-dimensional stochastic process, which is not necessarily ergodic.
Let $w$ be its winding number and $N_L$ (resp. $N_R$) be the number of its zeromode under the left (resp. right) SIBC.
Then, $w$ is equal to their difference; $w=N_L-N_R$.}

Since we assume ergodicity of the system, it suffices to show that the system has boundary-localized steady state under the left- (resp. right-) SIBC when the winding number is $w = 1$ (resp. $w=-1$).
We only consider the left-SIBC case with $w=1$ in the rest of this section.
The discussion below is valid in the right-SIBC case with $w=-1$ since there is an equivalence between the left and the right boundary $\bra{J_R(\beta)} = -\beta^L\bra{J_L(\beta)}$ (see the section above).
To prove the theorem, we utilize the non-Bloch wave expansion of the steady state
\begin{align}
\langle n|p_\textrm{ss}\rangle &= \sum_j c_j (\beta_j)^n |\phi_j(\beta_j)\rangle, \label{eqn: non-Bloch wave expansion} \\
W(\beta_j)\lvert\phi_j(\beta_j)\rangle &= 0. \label{eqn: bulk eigenequation}
\end{align}
In addition, we call the equation \eqref{eqn: left boundary equation} the (left) boundary equation in this section.

\subsubsection{Disappearance of the delocalized component}
\begin{figure}[t]
        \centering
        \includegraphics[width=140mm, bb=0 0 650 350, clip]{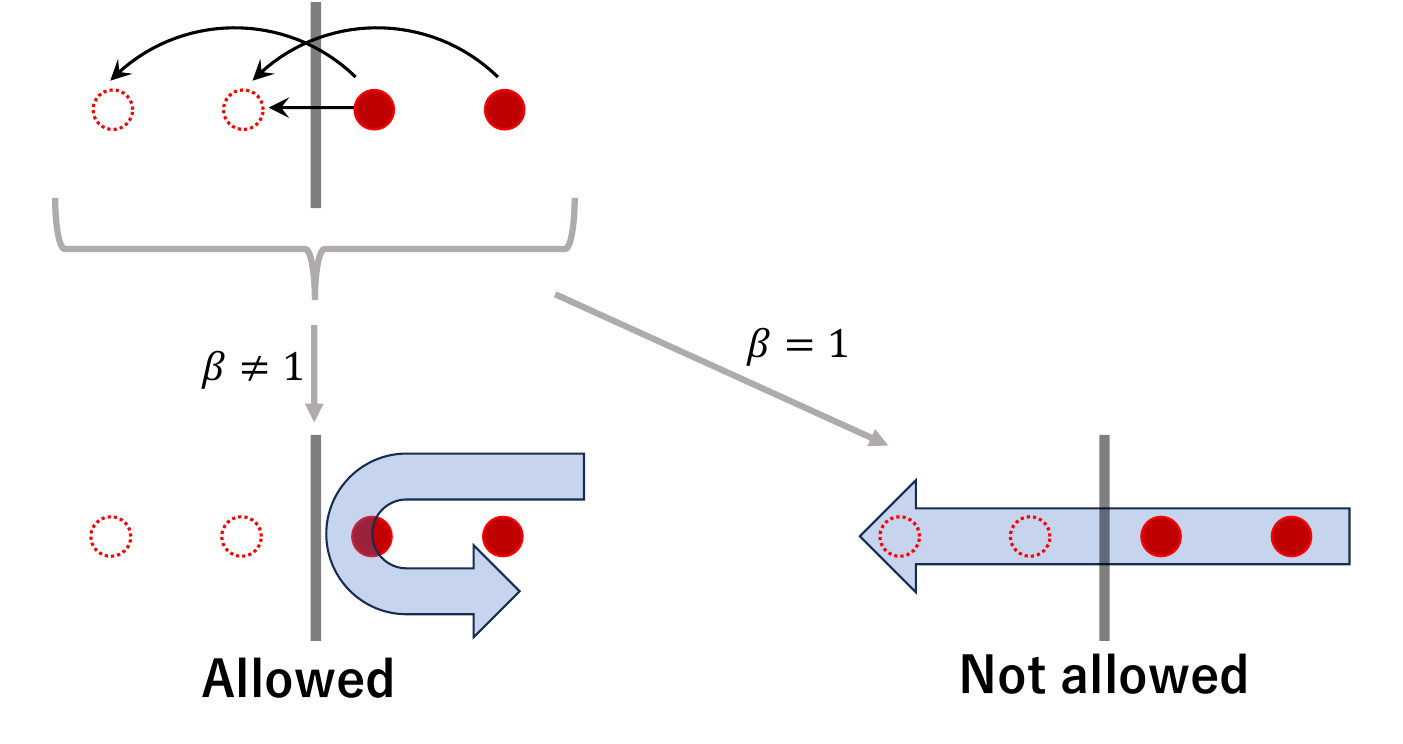}
        \caption{
        Schematic of the disappearance of the delocalized component of the steady state.
        The upper left figure represents the virtual current of the steady state.
        The lower left figure represents the allowed non-Bloch wave corresponding to $\beta \neq 1$.
        The right figure shows the prohibition of the delocalized non-Bloch wave corresponding to $\beta = 1$.
        These illustrate the mechanism of why steady state is localized when the winding number is nonzero.
        }
        \label{fig: Zerocurrent Condition}
    \end{figure}

We start with the proof of the current bulk-boundary correspondence $\bra{1}W(\beta) = \bra{J_L(\beta_{j})}$ by direct calculation:
  \begin{align}
    \bra{1}W(\beta)_\nu &= \sum_{\sigma}\sum_{m=-l_0}^{l_0} W_{m0;\sigma \nu}(\beta)^{-m} \notag \\
    &= \sum_{\sigma} \left(\sum_{m=-l_0}^{-1} + \sum_{m=1}^{l_0}\right)\left(W_{m0;\sigma \nu}(\beta)^{-m}-W_{-m0;\sigma \nu}\right) \notag \\
    &= \sum_{\sigma} \left(\sum_{m=1}^{l_0}\left(W_{m0;\sigma \nu}(\beta)^{-m}-W_{-m0;\sigma \nu}\right) + \sum_{m=1}^{l_0}\left(W_{-m0;\sigma \nu}(\beta)^{m}-W_{m0;\sigma \nu}\right)\right) \notag \\
    &= \sum_{\sigma} \left(\sum_{m=1}^{l_0}\left(W_{m0;\sigma \nu}(\beta)^{-m}-W_{-m0;\sigma \nu}\right) + \sum_{m=1}^{l_0}\left(W_{-m0;\sigma \nu}-W_{m0;\sigma \nu}(\beta)^{-m}\right)(\beta)^{m}\right) \notag \\
    &= \sum_{\sigma} \sum_{m=1}^{l_0}\left(W_{m0;\sigma \nu}(\beta)^{-m}-W_{-m0;\sigma \nu}\right)(1-\beta^m) \notag \\
    &= \beta^{-1}(1-\beta)\sum_{\sigma} \sum_{m=1}^{l_0}\left(W_{m0;\sigma \nu}(\beta)^{-m}-W_{-m0;\sigma \nu}\right)\left(\sum_{n=1}^m \beta^n\right) \notag \\
    &= \beta^{-1}(1-\beta)\bra{J_L(\beta)}.\label{eqn: proof of current bulk-boundary correspondence}
  \end{align}
Using the current bulk-boundary correspondence, we obtain the equality
    \begin{align}
        \braket{J_L(\beta_{j})}{\phi_{j}(\beta_{j})} = \lim_{\beta \to \beta_j}\frac{\bra{1}W(\beta)\ket{\phi_j(\beta)}}{(\beta^{-1} - 1)} 
        = \begin{cases}
                & 0 \: (\beta_j \neq 1),\\
                -&\left(\partial_\beta E(\beta)\right)_{\beta=1} \: (\beta_j = 1)
            \end{cases}
            \label{eqn: current from delocalization}
    \end{align}
for all $j$.
Then, the equation $\sum_j c_j \langle J_L(\beta_j) | \phi(\beta_j)\rangle = 0$ reads
    \begin{equation}
        c_{j_0} \left(\partial_\beta E(\beta)\right)_{\beta=1} = 0.
    \end{equation}
From the correspondence between the nonzero winding number $w$ and the nonzero first order derivative $\partial_\beta E$, we always obtain $c_j = 0$ when $w\neq 0$.
It completes the proof of the disappearance of the delocalization wave component in the steady state since delocalized component is only $\beta = 1$.
We give the schematic of the disappearance of the delocalized component in Fig.~\ref{fig: Zerocurrent Condition}.

\subsubsection{Localized steady state under the SIBC with a nonzero winding number}
To obtain the localized steady state, we use a localized non-Bloch waves ansatz:
\begin{align}
\langle n|p_\textrm{ss}\rangle &= \sum_{j=1}^{l_0 K} c_j (\beta_j)^n |\phi_j(\beta_j)\rangle, \label{eqn: localized steady state ansatz} \\
W(\beta_j)\lvert\phi_j(\beta_j)\rangle &= 0. \label{eqn: bulk eigenequation}
\end{align}
We note that the number of the solutions $\beta$ of $\det(W(\beta) = 0$ with $\abs{\beta} < 1$ is determined as $N^- = l_0K$ by the relationship between the winding number and the number of roots shown by residue theorem.
Then, Eq.~\eqref{eqn: left SIBC} reads
  \begin{align}
    &B^\textrm{Left}\bm{c} = 0 ,\\
    \end{align}
by using a $l_0K \times l_0K$ square matrix $B^\textrm{Left}$, whose components are described as
    \begin{align}
    &B^\textrm{Left}_{(n,\sigma), j} := \sum_\nu \sum_{m=n}^{l_0} \left(W_{m,0;\sigma \nu}\left(\beta_j\right)^{-m} - W_{-m,0;\sigma \nu}\right) \left(\beta_j\right)^n \phi(\beta_j)_\nu \\
    &(n= 1,\ldots,l_0, \sigma= 1, \ldots,q, j= 1,\ldots, l_0 K), \notag
  \end{align}
where $\bm{c}:=(c_j)_{j=1}^{l_0 K}$ is the coefficient vector.
We have 
  \begin{align}
    \sum_{n,\sigma} B^\textrm{Left}_{(n,\sigma), j} = \braket{J_L(\beta_j)}{\phi(\beta_j)} = 0
  \end{align}
for all $j$ by the definitions.
The last equality comes from the current bulk-boundary correspondence.
Therefore, $B^\textrm{Left}$ has the left eigenvector $(1,\ldots,1)$ of the zero eigenvalue, meaning that $B^\textrm{Left}$ must have nontrivial right eigenvector of zero eigenvalue.
Since the steady state satisfies the localized non-Bloch wave ansatz \eqref{eqn: localized steady state ansatz}, it is shown that the system has one left-localized steady state under the left SIBC when the winding number is one.

\subsubsection{Delocalized steady state under the SIBC with a zero winding number}
Delocalization is also discussed by the same tools.
The differences from nonzero-winding-number case are the equation $\langle J_L(\beta=1) | \phi(\beta=1)\rangle = 0$ and $N^\pm = lK_0 \pm 1$.
Therefore, in addition to the localized non-Bloch waves, we need the delocalized wave corresponding to $\beta = 1$ to construct the steady state.
This completes the proof of the delocalization of the steady state when the winding number is zero.
We note that if the rank of $B^\textrm{Left}$ is less than $lK_0-2$, the steady state may eventually be localized to the boundary.
However, such localization does not exhibit robustness against disorder that keeps the zero winding number unchanged.

\subsubsection{bulk-boundary correspondence under the OBC}
We clarify and prove the weaker statement of the theorem for the OBC.

\textbf{Proposition:} \textit{Consider a translationally invariant one-dimensional stochastic process, which is not necessarily ergodic.
Let $w$ be its winding number and assume $w\neq 0$.
Then, the system has a localized steady state under the OBC with finite system size $L$.}

In line with the proof in the SIBC case, we use the non-Bloch wave expansion \eqref{eqn: non-Bloch wave expansion}.
We obtain $c_{j_0} = 0$ by the arguments provided before; the current bulk-boundary correspondence and the equivalence of the left and the right boundaries.
Moreover, we also show that the steady state is written as localized wave expansion $\langle n|p_\textrm{ss}\rangle = \sum_{j\neq j_0} c_j (\beta_j)^n |\phi_j(\beta_j)\rangle$ by seeing the rank of the system of equations at the left and the right boundaries is $2l_0 K -1$.
The nonergodic case is immediately follows from ergodic case since all the lemmas in the main text are also holds true in the OBC case, which enables us to consider the bulk-boundary correspondence in each ergodic subsystems separately.

\subsection{Analytical calculations in the ergodic examples}
\begin{figure}[t]
        \centering
        \includegraphics[width=140mm, bb=0 0 900 200, clip]{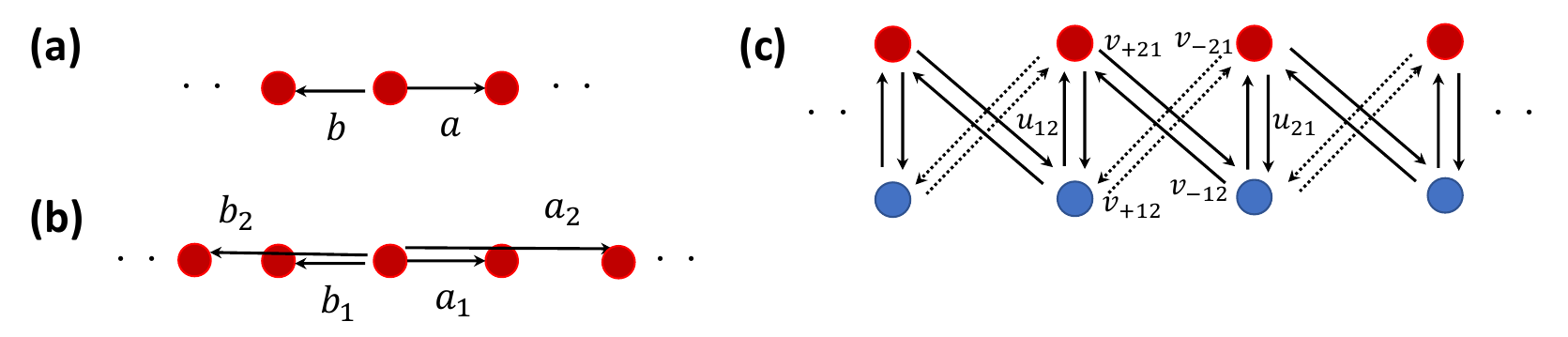}
        \caption{
        Transition diagrams of the models used in the numerical calculations.
        (a) The asymmetric random walk.
        (b) The 2-random walk.
        (c) The model with two internal degrees of freedom.
        }
        \label{fig: Transition diagrams theoretical example}
    \end{figure}

We demonstrate the current bulk-boundary correspondence in the examples shown in Fig.~\ref{fig: Transition diagrams theoretical example}.

\subsubsection{The asymmetric random walk}
The non-Bloch Hamiltonian of the asymmetric random walk is:
    \begin{equation}
        W(\beta) = a \beta^{-1} + b \beta -(a + b). \label{eqn: non-Bloch Hamiltonian asymmetric random walk}
    \end{equation}

The bulk equation and the system of equations at the boundaries are
  \begin{align}
    0 &= a p(n-1) + b p(n+1) - (a + b) p(n) \: (n = 2, \ldots, L-1), \label{eqn: asymmetric random walk bulk}\\
    0 &= b p(2) - a p(1) \: (n = 1), \label{eqn: asymmetric random walk left boundary} \\
    0 &= a p(L-1) - b p(L) \: (n = L). \label{eqn: asymmetric random walk right boundary}
  \end{align}

We write the two solution of the quadratic equation $\det(W(\beta)) = 0$ as $\beta_1=1, \beta_2 = a/b$.
We note that this notation is not consistent with the general case since the absolute value of $a/b$ is changed.
The non-Bloch wave expansion of the steady state is written as
  \begin{equation}
    p(n) = c_1 \beta_1^n + c_2 \beta_2^n.
  \end{equation}
Then it satisfies the bulk equation \eqref{eqn: asymmetric random walk bulk}.
By taking the difference between \eqref{eqn: asymmetric random walk left boundary},\eqref{eqn: asymmetric random walk right boundary} from the bulk equation, we obtain
  \begin{align}
    c_1 (a - b \beta_1) + c_2 (a - b \beta_2) &= 0, \\
    c_1 (b \beta_1^{L+1} - a \beta_1^{L}) + c_2 (b \beta_2^{L+1} - a\beta_2^{L}) &= 0.
  \end{align}
These equations give the explicit formulae of the left and the right boundary currents.
  \begin{align}
    J_L(\beta) &= a  - b \beta, \label{eqn: left boundary current asymmetric random walk}\\
    J_R(\beta) &= b \beta^{L+1} - a \beta^L. \label{eqn: right boundary current asymmetric random walk}
  \end{align}
By using equations \eqref{eqn: non-Bloch Hamiltonian asymmetric random walk}, \eqref{eqn: left boundary current asymmetric random walk}, and \eqref{eqn: right boundary current asymmetric random walk}, we can check the current bulk-boundary correspondence $\beta^{-1} (1-\beta) J_L(\beta) = W(\beta)$ and the equivalence of the boundary $J_L(\beta) = - \beta^{L} J_R(\beta)$.

  
\subsubsection{The $2$-random walk}
The non-Bloch Hamiltonian of the $2$-random walk is
  \begin{equation}
    W(\beta) = a_2 \beta^{-2} + a_1 \beta^{-1} + b_1 \beta + b_2 \beta^2 - (a_2 + a_1 + b_1 + b_2).
  \end{equation}

The system of equations at the boundaries are
  \begin{align}
    Ep(1) &= b_1 p(2) + b_2 p(3) - (a_1 + a_2)p(1), \\
    Ep(2) &= a_1 p(1) + b_1 p(3) + b_2 p(4) - (a_2 + a_1 + b_1)p(2), \\
    Ep(L-1) &= a_2 p(L-3) + a_1 p(L-2) + b_1 p(L) - (a_1 + b_1 + b_2)p(2), \\
    Ep(L) &= a_2 p(L-2) + a_1 p(L-1) - (b_1 + b_2)p(L).
  \end{align}
The upper (resp.) two equations are for the left (right) boundary.

We use the non-Bloch wave expansion $p(n) = \sum_j c_j (\beta_j)^n$ and take the difference of the system of equations at the boundaries from the bulk eigenvalue equation $E = a_2 (\beta_j)^{-2} + a_1 (\beta_j)^{-1} + b_1 (\beta_j) + b_2 (\beta_j)^2 - (a_2 + a_1 + b_1 + b_2)$ then we obtain
  \begin{align}
    0 &= \sum_j c_j \left[a_2 (\beta_j)^{-1} + a_1 - (b_1 + b_2)\beta_j\right] \\
    0 &= \sum_j c_j \left[a_2 - b_2(\beta_j)^2\right] \\
    0 &= \sum_j c_j \left[b_2 (\beta_j)^{L+1} - a_2(\beta_j)^{L-1}\right] \\
    0 &= \sum_j c_j \left[b_2 (\beta_j)^{L+2} + b_1 (\beta_j)^{L+1} - (a_2 + a_1)(\beta_j)^L\right]
  \end{align}
Therefore, the left and the right boundary currents are derived as
  \begin{align}
    J_L(\beta) &= a_2 \beta^{-1} + (a_2 + a_1) - (b_1 + b_2)\beta - b_2\beta^2, \\
    J_R(\beta) &= - a_2 \beta^{L-1} - (a_2 + a_1)\beta^L + (b_1 + b_2)\beta^{L+1} + b_2\beta^{L+2}
  \end{align}
and we show the current bulk-boundary correspondence $(\beta^{-1}-1)J_L(\beta) = W(\beta) $ and the equivalence of the boundaries as follows:
  \begin{align}
    (\beta^{-1}-1) J_L(\beta) &= a_2 \beta^{-2} + a_1 \beta^{-1} - (a_2 + a_1 + b_1 + b_2) + b_1\beta + b_2\beta^2 \\
    &=W(\beta), \\
    -\beta^L J_L(\beta) &= - a_2 \beta^{L-1} - (a_2 + a_1)\beta^L + (b_1 + b_2)\beta^{L+1} + b_2 \beta^{L+2} \\
    &= J_R(\beta).
  \end{align}
  
\subsubsection{A model with two internal degrees of freedom}
The non-Bloch Hamiltonian of the model with two internal degrees of freedom is
    \begin{align}
        W(\beta)
        = \begin{pmatrix}
            \tilde{d}_1 & u_{12} + v_{+12}\beta^{-1} + v_{-12}\beta \\
            u_{21} + v_{+21}\beta^{-1} + v_{-21}\beta & \tilde{d}_2
        \end{pmatrix}
    \end{align}
with $\tilde{d}_1$ and $\tilde{d}_2$ being $\tilde{d}_1 = -(u_{21} + v_{+21} + v_{-21})$, $\tilde{d}_2 = -(u_{12} + v_{+12} + v_{-12})$.

The eigenequations at the boundaries are
  \begin{align}
    E\bm{p}(1) &= \begin{pmatrix}
            0 & v_{-12} \\
            v_{-21} & 0
        \end{pmatrix} \bm{p}(2) 
         + \begin{pmatrix}
            -(u_{21} + v_{+21}) & u_{12} \\
            u_{21} & -(u_{12} + v_{+12})
        \end{pmatrix} \bm{p}(1), \\
    E\bm{p}(L) &= \begin{pmatrix}
            0 & v_{+12} \\
            v_{+21} & 0
        \end{pmatrix} \bm{p}(L-1) 
         + \begin{pmatrix}
            -(u_{21} + v_{-21}) & u_{12} \\
            u_{21} & -(u_{12} + v_{-12})
        \end{pmatrix} \bm{p}(L).
  \end{align}
The upper (lower) equation is for the left (right) boundary.

We use the non-Bloch wave expansion $\bm{p}(n) = \sum_j c_j (\beta_j)^n \ket{\phi(\beta_j)}$ where $\ket{\phi(\beta_j)}$ is explicitly written as:
    \begin{align}
        \ket{\phi(\beta_j)} = \frac{1}{\mathcal{N}(\beta_j)}\begin{pmatrix}
            u_{12} + v_{+12}\beta_j^{-1} + v_{-12}\beta_j  \\
            u_{21} + v_{+21} + v_{-21}
        \end{pmatrix}.
    \end{align}
$\mathcal{N}(\beta_j) = (u_{21} + v_{+21} + v_{-21}) + (u_{12} + v_{+12}\beta_j^{-1} + v_{-12}\beta_j)$ is the normalization constant corresponding to the condition $\braket{1}{\phi(\beta_j)} = 1$.

The differences between the eigenequations at the boundaries and the bulk eigenvalue equation read
  \begin{align}
    0 &= \sum_j c_j \left[\beta_j\begin{pmatrix}
            v_{-21} & -v_{+12}\beta_j^{-1} &
            -v_{+21}\beta_j^{-1} & v_{-12}
        \end{pmatrix} \ket{\phi(\beta_j)} \right], \\
    0 &= \sum_j c_j \left[\beta_j^{L}\begin{pmatrix}
            v_{+21} & -v_{-12}\beta_j &
            -v_{-21}\beta_j & v_{+12}
        \end{pmatrix} \ket{\phi(\beta_j)} \right].
  \end{align}
Summation with respect to internal degrees of freedom, i.e., the multiplication of $\bra{1}$ gives
    \begin{align}
        0 &= \sum_j c_j \left[\beta_j\begin{pmatrix}
            v_{-21}-v_{+21}\beta_j^{-1}&
            v_{-12}-v_{+12}\beta_j^{-1}
        \end{pmatrix} \ket{\phi(\beta_j)} \right], \\
        0 &= \sum_j c_j \left[\beta_j^{L}\begin{pmatrix}
            v_{+21} - v_{-21}\beta_j &
            v_{+12} - v_{-12}\beta_j
        \end{pmatrix} \ket{\phi(\beta_j)} \right]
    \end{align}
Therefore, we obtain the explicit formulae of the left and the right boundary currents
  \begin{align}
    \bra{J_L(\beta_j)} &= \beta_j\begin{pmatrix}
            v_{-21}-v_{+21}\beta_j^{-1}&
            v_{-12}-v_{+12}\beta_j^{-1}
        \end{pmatrix} 
        = -\begin{pmatrix}
            v_{+21} - v_{-21}\beta_j &
            v_{+12} - v_{-12}\beta_j
        \end{pmatrix}, \\
    \bra{J_R(\beta_j)} &= \beta_j^{L}\begin{pmatrix}
            v_{+21} - v_{-21}\beta_j &
            v_{+12} - v_{-12}\beta_j
        \end{pmatrix}.
  \end{align}
We can straightforwardly check the current bulk-boundary correspondence $(\beta^{-1}-1)J_L(\beta) = \bra{1}W(\beta) $ and the equivalence of the boundaries $\bra{J_L(\beta)} = - \beta^{L} \bra{J_R(\beta)}$.

\subsection{Comparison to the previous research}
We comment on the difference between this Letter and the previous research \cite{Murugan2017} that addresses the localization of a steady state using Hermitianization.
The previous research has focused on the case where the system is a composite of two systems with infinite length and in topologically nontrivial phases.
Therefore, it cannot be applied to the case where one of the subsystems is the vacuum or in the topologically trivial phase.
In contrast, we prove the localization at the boundary between the system and the vacuum and demonstrates the localization even in finite systems.

\subsection{Proof of the lemmas in nonergodic case}
We give the proof of the following Lemmas 1, 2 and 3.
Below $W$ and $w$ denotes a transition matrix that is not necessarily an ergodic stochastic process and its winding number.

\textbf{Lemma 1: } \textit{The number of zeromodes of $W$ increases in $O(L)$ with $L$ being the system size, or the sink components of the graph induced by $W$ also have translation invariance.}

\textbf{Lemma 2: } \textit{The winding number $w$ is the sum of each winding number $w_i$ calculated separately from each sink component.}

\textbf{Lemma 3: } \textit{The steady states have nonzero amplitude only on sites in the sink components.}




For the sake of completeness, we give the precise definitions of graph-theoretic concepts used in the lemmas: induced graph, strongly connected component, and symmetry.
We often write a directed graph $G$ as $G=(V,E)$ where $V$ is the set of vertices of $G$ and $E \subset V\times V$ is the set of directed edges of $G$.

\textbf{Definition 1: } \textit{We say that a directed graph $G$ is induced by a matrix $W$ when the connection matrix of $G$ is obtained from the replacement of the nonzero off-diagonal components of $W$ by $1$ and others by $0$.}

\textbf{Definition 2: } \textit{We say that a directed graph $G$ is strongly connected if and only if for any two vertices $i$ and $j$, there is a path from $i$ to $j$. Moreover, the subgraph $G_s$ of $G$ is called strongly connected component when the $G_s$ is strongly connected.}

\textbf{Definition 3: } \textit{Let $G=(V,E)$ is a directed graph. We say that a map $T: V \to V$ is a symmetry of $G$ when $T$ has the following property: $(i,j) \in E$ if and only if $(T(i),T(j)) \in E$.}

\subsubsection{Lemma 4}
\begin{figure}[t]
        \centering
        \includegraphics[width=85mm, bb=0 0 300 230, clip]{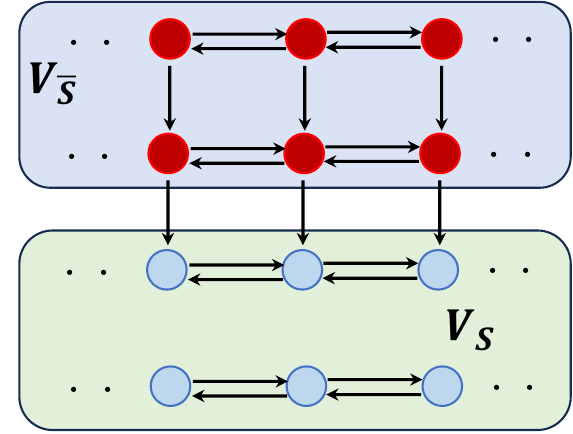}
        \caption{
        Illustration of the definition of $V_S$ and $V_{\bar{S}}$in the Lemma 4.
        The sites in $V_s$ have edge only directed to the strongly connected component they belong to.
        }
        \label{fig: Lemma decomposition of strongly connected component}
    \end{figure}

We utilize another following lemma.
The proof of the Lemma 4 is given after that of the Lemmas 1,2 and 3.
We provide the illustration of $V_S$ and $V_{\bar{S}}$ in Fig.~\ref{fig: Lemma decomposition of strongly connected component}.

\textbf{Lemma 4: } \textit{Let $W$ and $G = (V,E)$ be the transition-rate matrix and the corresponding directed graph of a nonergodic process.
We write the union set of the vertices of all sink components as $V_S$ and define $V_{\bar{S}} := V \setminus V_S$.
Then, as for the block-matrix representation of $W$ with respect to $V_{\bar{S}}, V_S$
  \begin{align}
    W = \begin{pmatrix}
        A_{\bar{S}} & 0 \\ W_{S\bar{S}} & W_S
    \end{pmatrix},
  \end{align}
all the real parts of the eigenvalues of $A_{\bar{S}}$ are smaller than zero.
Furthermore, any steady state of $W$ takes zero value at $V_{\bar{S}}$.}

Since the Lemma 3 is included as the special case of the Lemma 4, we only prove the Lemmas 1 and 2.
    
\subsubsection{Proof of the Lemma 1}
The Lemma 1 is proved by a simple argument below.
We consider a directed graph $G$ and its general symmetry $T: G \to G$.
Let $G_i$ ($i=1,2,\ldots$) denote the strongly connected components of $G$.
Then, for all $i$, there exists $j$ such that $T(G_i) = G_j$ since the symmetry $T$ preserves the strong connectivity.
Therefore, for fixed strongly connected component $G_i$, one of the following cases always holds true: 
(1) There exists some integer $m(i)$ such that $T^{m(i)}(G_i) = G_i$. 
(2) For all integer $m$, $T^m(G_i) \neq G_i$.

In the case of (1), $G_i$ has $m(i)$-multiple translation invariance of the original system $G$.
In the case of (2), the number of strongly connected components of $G$ is $O(L)$ with $L$ being system size.
The former of the statement of the Lemma 1 is the case that there is a $G_i$ that satisfy the case (2) and the steady state of $W$ can be nonzero at sites of $G_i$, since the number of steady states is larger than the number of isolated strongly connected components isomorphic to $G_i$, which is $O(L)$.
In the other case, since all $G_i$ that satisfy (2) are not sink components, all sink components satisfy (1).
Therefore, all the sink component $G_i$ has translation invariance with operator $T^{m(i)}$.
This is the latter statement of the Lemma 1.

\subsubsection{Proof of the Lemma 2}
Let $G$ be the induced graph of $W$.
We write each strongly connected component of $G$ as $G_i$.
We only consider the case that each sink component has translation invariance so that the winding number is well-defined.

Firstly, we note that the winding number is determined by the contribution of the sink components.
This is shown by representing the winding number as the contribution of the diagonal block matrix in the Jordan normal form of $W$ with respect to strongly connected components.
The Lemma 4 tells us that the extra diagonal loss makes all the spectra have negative real parts.
In the following, we only consider the case where $G$ is composed of sink components $G_i$ ($i=1, \ldots, N_c$) where $N_c$ is a finite integer independent of the system size $L$.
We further assume $T(G_i) = G_{i+1 \mod N_c}$ with T being the translation operator.
This assumption indicates that all the strongly connected components are equivalent by the translation and it immediately follows that each $G_i$ has the translation invariance $T^{N_c}$.
In this sense, we can say that $G$ has a single cycle of sink components $G_1, \ldots, G_{N_c}$ with respect to translation.
The discussion given below is straightforwardly extended to the case where $G$ has more than one cycle with respect to the translation.

We consider the periodic system with the system size $L= N_c L_0$ where $L_0$ is an integer.
The Bloch ansatz under this PBC takes the form of
  \begin{equation}
        \braket{n}{\psi(k)} = e^{ikn} \ket{\psi(k)},
  \end{equation}
where the wavenumber $k$ takes the value of $0, \ldots, \frac{L-1}{L}2\pi$.
The dimension of the corresponding Bloch Hamiltonian $W(k)$ is equal to the number of internal degrees of freedom.

Then, we replace the unit cell with the $N_c$-multiple of it.
Precisely, we define an extended unit cell by regarding unit cells labeled with $n = (l-1)N_c + 1, (l-1)N_c+2, \ldots, (l-1)N_c + (N_c - 1)$ as one unit cell labeled with $l$.
We call such an extended unit cell $N_c$-unit cell.
When it comes to the Bloch ansatz $\braket{n}{\psi(k)} = e^{ikn} \ket{\psi(k)}$ with respect to the $N_c$-cells, the range of the wavenumber $k$ becomes
  \begin{align}
    k = 0, \ldots, \frac{L_0-1}{L_0} \frac{2\pi}{N_c},
  \end{align}
since the factor $e^{ikn}$ gains $e^{ikN_c}$ by the shift of the $N_c$-unit cell.
We note that for the Bloch Hamiltonian $\widetilde{W}^{N_c}(k)$ with respect to $N_c$-unit cell, the wavenumber appears only in the form of $e^{ikN_c}$.

Since each $G_i$ has translation invariance $T^{N_c}$, $\widetilde{W}^{N_c}(k)$ can be written as the block diagonal matrix with diagonal entries $W_i$ under an appropriate basis.
The crucial point is that the wavenumber $\tilde{k}$ used in the Bloch Hamiltonian $W_i(\tilde{k})$ of each $W_i$ is $N_c$-multiple of the $k$ which is used in the original $W(k)$ since $\tilde{k}$ corresponds to $T^{N_c}$, i.e., $\tilde{k} = N_c k$.
Therefore, we obtain the matrix relation
  \begin{align}
    \widetilde{W}^{N_c}(k) \propto
    \begin{pmatrix}
        W_1(N_ck) &  & & \\ & W_2(N_ck) & & \\ & & \ddots & \\ & & & W_M(N_ck),
    \end{pmatrix}
  \end{align}
which shows the separation of the spectrum of $W(k)$ to those of $W_i(N_c k)$ ($i= 1, \ldots N_c$).

By taking the limit $L_0 \to \infty$ with $N_c$ fixed, we prove that we can calculate the winding number $w_\pm$ of $W$ as the contribution of those of $W_i$:
  \begin{align}
    w^\lambda = \int_0^{2\pi/N_c} \frac{1}{2\pi i} \textrm{d}k \frac{\partial}{\partial k} \log(\det(\widetilde{W}^{N_c}(k + i\lambda))) 
    &= \sum_j \int_0^{2\pi/N_c} \frac{1}{2\pi i} \textrm{d}k \frac{\partial}{\partial k} \log(\det(W_j(N_ck + iN_c\lambda))) \notag \\
    &= \sum_j \int_0^{2\pi} \frac{1}{2\pi i} \textrm{d}k \frac{\partial}{\partial k} \log(\det(\tilde{W_j}(k + iN_c\lambda))) \notag \\
    &= \sum_j w_j^{N_c\lambda}, \\
    w_\pm &= \lim_{\lambda \to \pm 0} w^\lambda \notag \\
    &= \sum_j w_{j,\pm}.
  \end{align}
It completes the proof of the Lemma 2.

\subsubsection{Proof of the Lemma 4}
We show the Lemma 4 in the previous subsection.
The latter part immediately follows from the former part by the eigenvalue equation $W\ket{p_{\textrm{ss}}} = 0$.
We prove the former part of the statement.

Initially, we express $A_{\bar{S}} = W_{\bar{S}} - D_{\bar{S}}$, with $W_{\bar{S}}$ representing the transition-rate matrix corresponding to $G_{\bar{S}}$ and 
    \begin{align}
        D_{\bar{S}} = \textrm{diag}\left(\sum_j (W_{S\bar{S}})_{1j},\cdots,\sum_j (W_{S\bar{S}})_{N_{\bar{S}}j}\right)
    \end{align}
being the block diagonal matrix determined by the off-diagonal block matrix $W_{S\bar{S}}$.
It is worth noting that all the diagonal elements of $D_{\bar{S}}$ are nonnegative, and for all the strongly connected components in $G_{\bar{S}}$, there is a vertex $i$ such that $(D_{\bar{S}})_{ii} > 0$.
Since $A_{\bar{S}}$ can be regarded as a nonnegative matrix shifted by a real negative constant, the existence of a real eigenvalue $\Lambda_A$ of $A_{\bar{S}}$ and the following properties are derived by the Perron-Frobenius theorem.
Firstly, there must exist an eigenvector $\ket{\psi_0}$ of $A_{\bar{S}}\ket{\psi_0} = \Lambda_A\ket{\psi_0}$ with all the components of $\ket{\psi_0}$ being nonnegative.
Furthermore, any eigenvalue $\Lambda$ of $A_{\bar{S}}$ except $\Lambda_A$ satisfies $\textrm{Re} \Lambda < \Lambda_A$.
Based on the above considerations, it suffices to show $\Lambda_A < 0$.
Utilizing $\bra{1} W = 0$, we obtain the formula for $\Lambda_A$ as follows:
  \begin{align}
    \bra{1}A_{\bar{S}}\ket{\psi_0} &= \Lambda_A\braket{1}{\psi_0} = - \bra{1} D_{\bar{S}} \ket{\psi_0}, \\
    \Lambda_A = - \frac{\bra{1} D_{\bar{S}} \ket{\psi_0}}{\braket{1}{\psi_0}}.
  \end{align}
This implies $\Lambda_A \leq 0$ from inequalities $\braket{1}{\psi_0}>0$ and $\bra{1}D_{\bar{S}} \ket{\psi_0} = \sum_i (D_{\bar{S}})_{ii} (\psi_0)_i \geq 0$ with $(\psi_0)_i$ being the $i$th component of $\ket{\psi_0}$.

We further prove $\Lambda_A \neq 0$ by contradiction.
We first assume $\Lambda_A = 0$.
Then, by the definition of $G_{\bar{S}}$, for each strongly connected components $G_j$, there must be an index $i_0$ such that $(D_{\bar{S}})_{i_0i_0} \neq 0$.
Therefore, for such $i_0$ we obtain $(\psi_0)_{i_0} = 0$ and it implies  $(\psi_0)_i = 0$ for any site $i$ in $G_j$ as we see in the next paragraph.
Therefore, $\ket{\psi_0} = 0$ is derived, which means the absence of the eigenvector $|\psi_0\rangle$ corresponding to $\Lambda_A = 0$.
That is a contradiction and completes the proof.

Before closing this subsection, we show that $(\psi_0)_{i_0} = 0$ implies $(\psi_0)_i = 0$ for any site $i$ in $G_j$.
Substituting $(\psi_0)_{i_0} = 0$ into the eigenvalue equation $((A_{\bar{S}} \ket{\psi_0})_{i_0} = 0$ at site $i_0$, we obtain
  \begin{align}
    \sum_{j:(j \to i) \in E_{\bar{S}}} (A_{\bar{S}})_{i_0j} (\psi_0)_j = 0,
  \end{align}
and we derive $(\psi_0)_j = 0$ since $(A_{\bar{S}})_{i_0j} > 0$.
Iterating this argument by replacing $i_0$ with $j$ we obtain $(\psi_0)_j = 0$, and thus it is shown that all the vertices $j_m$ which has a directed path to $i$ satisfies $(\psi_0)_{j_m} = 0$.
Therefore, $\ket{\psi_0}$ is always zero on the connected component $G_j$ because any pair of two vertices mutually have a directed path to each other.

\subsection{Bulk-boundary correspondence in the asymmetric simple exclusion process (ASEP)}
We define and calculate the winding number of the asymmetric simple exclusion process (ASEP) under the PBC.
Since the particle number $N$ is the conserved quantity under the PBC, we define the winding number for each subspace which is determined by $N$.

\subsubsection{Setup of the ASEP}
The master equation of the ASEP is:
  \begin{equation}
    \frac{\textrm{d}}{\textrm{d}t} P(x_1, \ldots, x_N) = \sum_{i} \left(a P(x_1, \ldots, x_i-1, \ldots, x_N) + b P(x_1, \ldots, x_i + 1, \ldots, x_N) -(a + b) P(x_1, \ldots, x_N) \right), \label{eqn: ASEP master equation}
  \end{equation}
where $P(x_1, \ldots, x_N)$ is the joint probability distribution of that the $i$th particle occupies the position $x_i$.
We impose the hardcore interaction on the ASEP,
    \begin{equation}
        P(x_1, \ldots, x_N) = 0 \: \textrm{if} \: x_i = x_j,
    \end{equation}
and modify the diagonal loss $-(a + b)$ in Eq.~\eqref{eqn: ASEP master equation} to compensate the absence of the hopping to forbidden configurations $\{ (x_1, \ldots, x_N) | x_i = x_j \: \textrm{for some} \: i \neq j.\}$.

We denote the transition-rate matrix of the $N$-particle ASEP as $M$ in this section.
Precisely, the off-diagonal components $M(\bm{x}, \bm{\tilde{x}})$ ($x_i \neq x_j$ and $\tilde{x}_i \neq \tilde{x}_j$ for all $i\neq j$) are determined as
    \begin{equation}
        \bm{x} \neq  \bm{\tilde{x}} \Rightarrow M(\bm{x}, \bm{\tilde{x}}) = \sum_j \left(a \delta(\bm{\tilde{x}}, \bm{x} - \bm{e}_j) + b \delta(\bm{\tilde{x}}, \bm{x} + \bm{e}_j)\right)
    \end{equation}
where $\bm{e}_j$ is the unit vector whose $j$th component is one and others are zero.
We note that the diagonal components $M(\bm{x},\bm{x})$ ($x_i \neq x_j$ for all $i\neq j$) are determined by the constraint $\sum_{\bm{\tilde{x}}} M(\bm{\tilde{x}},\bm{x}) = 0$.

\subsubsection{Definition of the winding number}
The ASEP has homogeneous translation invariance described as
  \begin{align}
    M(\bm{x}, \bm{\tilde{x}}) = M(\bm{x} + \bm{1}, \bm{\tilde{x} + \bm{1}})
  \end{align}
where $\bm{1}$ is a vector whose components are all equal to one.
This property means that the invariance under the uniform shift.
We note that the homogeneous translation invariance is not the same as the translation invariance which appear in the derivation of the Bethe equation.

We consider the Fourier transformation with respect to the homogeneous translation invariance and the corresponding Bloch Hamiltonian:
  \begin{align}
    P(\bm{x}) &\mapsto P(k;\bm{x}) \: \textrm{s.t.} \: P(k;\bm{x}+\bm{1}) = e^{ik}P(k;\bm{x}) \\
    M(\bm{x}, \bm{\tilde{x}}) &\mapsto M(k;\bm{x}, \bm{\tilde{x}})
  \end{align}
The explicit formula of $M(k;\bm{x}, \bm{\tilde{x}})$ cannot be obtained unless we specify the details of the Fourier transformation.
Using the Fourier transformation, we can construct the Brillouin zone $\{k = 2\pi n/L | n = 0, \ldots, L-1\}$ for the system with the size $L$.

We now define the winding number in a similar manner to the one-particle case.
We consider the imaginary gauge transformation $M^\lambda$ and calculate the limit $w_\pm := \lim_{\lambda \to +0} w(M^{\pm\lambda})$.
Finally, we define the winding number $w$ as $w=w_+ + w_-$.

\subsubsection{Calculation of the winding number using Bethe ansatz}
Since the eigenequation of the ASEP is exactly solved by using the Bethe ansatz,
  \begin{align}
    P(\bm{x}) = P_{\bm{z}}(\bm{x}) = \sum_{\sigma \in \mathfrak{S}} A_\sigma \prod_{j}z_{\sigma(j)}^{x_j},
  \end{align}
we can directly calculate the winding number.
Firstly, we note that the relationship between the Bethe roots and the wavenumber $k$ which appear from the homogeneous translation invariance.
Acting the homogeneous translation operation on the Bethe eigenfunction, we obtain the equality
  \begin{align}
    e^{ik} P_{\bm{z}}(k;\bm{x}) &= \sum_{\sigma \in \mathfrak{S}} A_\sigma \prod_{j} z_{\sigma(j)}^{x_j + 1} \label{eqn: Fourier transform of vector}\\
    &= \left(\prod_{j} z_{j} \right) \sum_{\sigma \in \mathfrak{S}} A_\sigma \prod_{j} z_{\sigma(j)}^{x_j} \\
    &= \left(\prod_{j} z_{j} \right) P_{\bm{z}}(k;\bm{x}).
  \end{align}
Therefore, we obtain the value of the first order derivative:
    \begin{equation}
        1 = \sum_{j} \partial_{(ik)}\vert_{k=0} (z_{j}).
    \end{equation}
Since the above calculation works for all the possible Fourier transformation \eqref{eqn: Fourier transform of vector}, the value of the winding number obtained below is not dependent on the details of the Fourier transformation.

Then we can calculate the first order derivative of the eigenvalue of the ASEP.
  \begin{align}
    \partial_{ik}\vert_{k=0} E_0(k) &= \sum_j (a \partial_{ik}\vert_{k=0}(z_{j}^{-1}) + b\partial_{ik}\vert_{k=0}(z_{j})) = \sum_j (b-a) \partial_{(ik)}\vert_{k=0} (z_{j})\\
    &= (b-a)
  \end{align}
where we used $\partial_{ik}\vert_{k=0}(z_{j}^{-1}) = -z_{j}(k=0)^{-2}\partial_{ik}\vert_{k=0}(z_{j})$ and $z_j(k=0) = 1$.
Based on the correspondence between the first-order derivative and the winding number, the last expression tells us the value of the winding number since the ASEP is ergodic.
We obtain $w = 1$, $0$, $-1$ corresponding to $a < b$, $a=b$, $a>b$.
This result is the extension of the asymmetric random walk, which can be regarded as the one-particle ASEP.

\subsubsection{Relationship between our winding number and the many-body winding number in a previous research}
Before ending this section, we comment on the relationship between our winding number and the many-body winding number introduced in the previous research \cite{PhysRevB.105.165137}.

We write the eigenvalues of the system under the twisted boundary condition $M_\theta(\bm{x},\bm{\tilde{x}} ) := M(\bm{x},\bm{\tilde{x}}) \prod_j e^{i\frac{\theta}{L}(x_j - \tilde{x}_j)}$ as 
  \begin{align}
    E_\theta(\bm{z}) &= \sum_i (ae^{-i\theta/L}z_i^{-1} + be^{i\theta/L}z_i - (a+b)) \\
    &=: \sum_i (a\tilde{z}_i(\theta)^{-1} + b\tilde{z}_i(\theta) - (a+b))
  \end{align}
by utilizing the Bethe ansatz.
This tells us that the twisted boundary condition is equivalent to the transformation of the Bethe root:
  \begin{align}
    z_{i} \mapsto \tilde{z}_i(\theta) = e^{i\theta /L} \tilde{z}_i.
  \end{align}
Since the system has homogeneous translation invariance for arbitrary $\theta$, we obtain 
  \begin{align}
    \det\left(M_\theta -E_B\right) &= \prod_{k^\prime \in \textrm{BZ}}\det\left(M\left(k^\prime -\theta \frac{N}{L}\right) - E_B\right)
  \end{align}
with $\textrm{BZ} = \left\{\frac{2\pi}{L} j \vert j = 0,1,\ldots,L-1 \right\}$ by the block diagonalization with respect to the Bloch Hamiltonian
  \begin{align}
    M_\theta(k;\bm{x},\bm{\tilde{x}}) &:= M(\bm{x},\bm{\tilde{x}}) \prod_j e^{i\frac{\theta}{L}(x_j - \tilde{x}_j)}e^{-ik\frac{1}{N}(x_j - \tilde{x}_j)} \\
    &= M\left(k-\theta \frac{N}{L};\bm{x},\bm{\tilde{x}}\right).
  \end{align}
We now prove the equivalence between our winding number and the previous one \cite{PhysRevB.105.165137} as below.
The calculation is the generalization of the proof of the equivalence in the one-particle case provided in the previous research \cite{PhysRevB.105.165137}.
  \begin{align}
    w_N(E_B) &= \sum_{k^\prime \in \textrm{BZ}} \int_0^{2\pi} \frac{\textrm{d}\theta}{2\pi i}\frac{\textrm{d}}{\textrm{d}\theta} \log{\det\left(M\left(k^\prime -\theta \frac{N}{L}\right) - E_B\right)} \\
    &= \sum_{k^\prime \in \textrm{BZ}} \int_{k^\prime}^{k^\prime - 2\pi \frac{N}{L}} \frac{\textrm{d}k}{2\pi i}\frac{\textrm{d}}{\textrm{d}k} \log{\det(M(k) - E_B)} \\
    &= - N\int_{0}^{2\pi} \frac{\textrm{d}k}{2\pi i}\frac{\textrm{d}}{\textrm{d}k} \log{\det(M(k) - E_B)}.
  \end{align}
Here, we calculated the range of the integration as:
  \begin{align}
    \sum_{k^\prime \in \textrm{BZ}} \int_{k^\prime}^{k^\prime - 2\pi N/L} &= -\sum_{k^\prime \in \textrm{BZ}} \sum_{l = 1,\ldots,N} \int_{k^\prime - 2\pi l/L}^{k^\prime - 2\pi (l-1)/L}\\
    &= -\sum_{j= 0, \ldots,L-1} \sum_{l = 1,\ldots,N} \int_{\frac{2\pi}{L} (j-l)}^{\frac{2\pi}{L}(j-l+1)} \\ 
    &= -\sum_{l = 1,\ldots,N} \sum_{j^\prime = 0, \ldots,L-1} \int_{\frac{2\pi}{L} j^\prime}^{\frac{2\pi}{L}(j^\prime +1)} \:\: (j^\prime := j-l) \\
    &= -N \int_{0}^{2\pi}.
  \end{align}

\subsubsection{Localized steady states and the bulk-boundary correspondence in the ASEP}
We confirm the correspondence between the localization of the steady state and the winding number by using the Bethe ansatz \cite{EPL.26.7,PhysRevLett.95.240601}.
When the system is under the reflective boundary condition, namely, the OBC with the particle number conservation, the steady state is rigorously obtained in each particle-number sector.
The paper \cite{EPL.26.7} provides the formula of the $N$-particle steady state $P_\textrm{ss}(\bm{x}) \propto q^{2\sum_j x_j}$ where $q = \sqrt{b/a}$ in our notation.
Therefore, the spatial asymmetry of the ASEP corresponds to the localization direction of the steady state under the OBC.
We note that one can obtain the steady state in another way of using the unitary transformation provided in Ref.~ \cite{PhysRevLett.95.240601}, which maps the ASEP to the XXZ model and calculates the multiplication of all the Bethe roots of the steady states of the XXZ model.

Even when the system lacks the particle-number conservation because of the existence of boundary injections (left: $\alpha$, right: $\gamma$) or ejections (left: $\beta$, right: $\delta$), the localization can be argued by the matrix Bethe ansatz given in \cite{JPhysMathGen.26.1493}.
The matrix Bethe ansatz are used on the probability $P(\tau_1, \ldots, \tau_L)$ of the configuration $(\tau_j)_{j=1}^L$ where $\tau_j$ is one when the particle exists at the site $i$ and zero otherwise.
When boundary terms are absent, the consistency of notations are given as $P(x_1, \ldots, x_N) = P(\tau_1, \ldots, \tau_L)$ when $\tau_{x_i} = 1$ ($i=1, \ldots, N$) and the other $\tau_j$'s are zero.
The explicit formulas of the matrix Bethe ansatz are follows:
    \begin{align}
        P(\tau_1, \ldots, \tau_L) &= \frac{1}{Z_L}\langle\langle W | \prod_i (\tau_i D + (1-\tau_i)E) | V\rangle\rangle, \\
        aDE - bED &= D+E, \\
        \langle\langle W |(\gamma D - \alpha E) &= \langle\langle W |, \\
        (\beta D - \delta E)| V\rangle\rangle &= | V\rangle\rangle,
    \end{align}
with $Z_L$ being a normalizing constant where $D$, $E$ are matrix analogy of Bethe roots and $\langle\langle W |$, $| V\rangle\rangle$ are auxiliary vectors to calculate the probability amplitude.
It is natural to expect that our bulk-boundary correspondence can be extended to nonconservative many-body systems by discussing the bulk winding number under the nonconservative OBC, which still remains unestablished even in general non-Hermitian systems.

\subsection{Further numerical results}
We give more detailed numerical results including the calculation of the winding number.
We calculate the winding number under the PBC by calculating $w^\lambda := (2\pi i)^{-1}\int_0^{2\pi} \frac{d}{dk}\log(\det(W^\lambda(k)))\mathrm{d}k$ at $\lambda = \lambda_+, -\lambda_-$, with $\lambda_\pm$ being sufficiently small constants, $\abs{\lambda_\pm} \ll 1$
In the calculation of the steady state under the OBC, we add the off-diagonal disorder discussed in the main text $W_{nm;\sigma\nu} \mapsto \tilde{W}_{nm;\sigma\nu} + \Delta_{nm;\sigma\nu}$ ($(n,\sigma)\neq (m,\nu)$) where each $\Delta_{nm;\sigma\nu}$ is randomly generated from the uniform distribution on $[-\delta, \delta]$ ($\delta > 0$).
We calculate the mean and the variance of the logarithm of the probability $p(n,\sigma)$ in 1000 samples to see the robustness of the localization.
We take the logarithm to obtain a normally distributed histogram.

\begin{figure}[t]
        \centering
        \includegraphics[width=140mm, bb=0 0 800 250, clip]{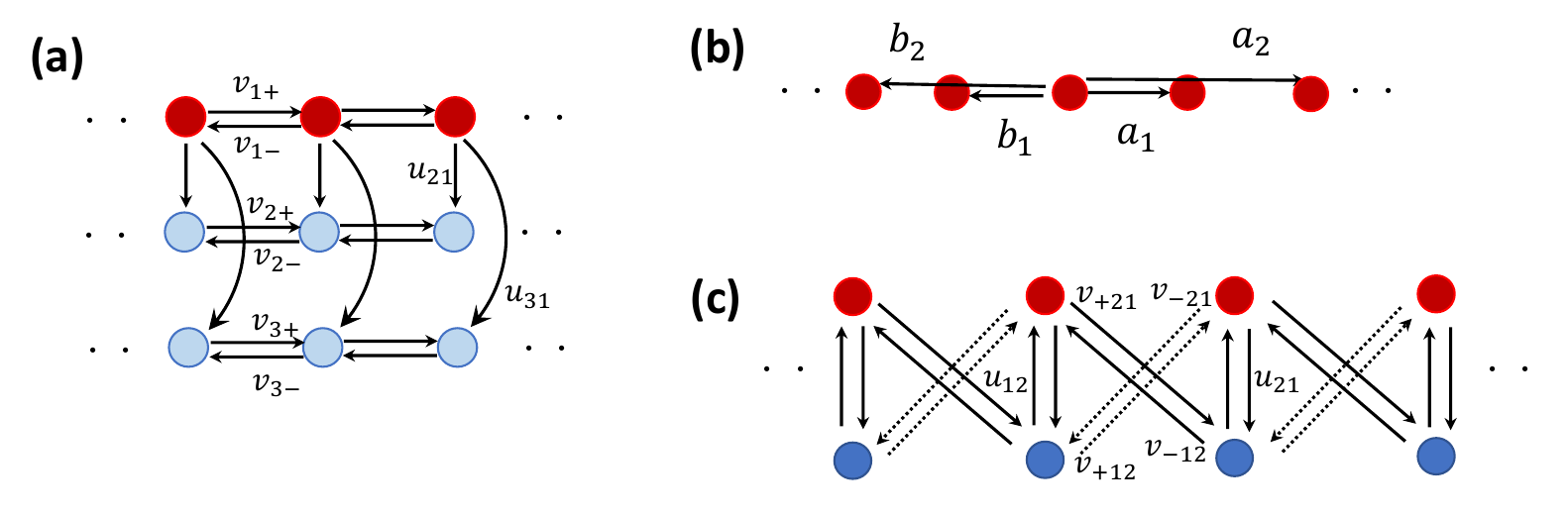}
        \caption{
        Transition diagrams of the models used in the numerical calculations.
        (a) The nonergodic model in the main text.
        (b) The 2-random walk.
        (c) The model with two internal degrees of freedom.
        }
        \label{fig: Transition diagrams}
    \end{figure}

We provide the transition diagrams of the models used in the numerical calculation (Fig.~\ref{fig: Transition diagrams}).
Figure~\ref{fig: Transition diagrams}(a) represents the nonergodic model referred as Eq.~(2) in the main text:
    \begin{align}
        \frac{\mathrm{d}}{\mathrm{d}t} \bm{p}(n,t)
        = \begin{pmatrix}
            v_{1+} & 0 & 0 \\
            0 & v_{2+} & 0 \\
            0 & 0& v_{3+}
        \end{pmatrix}\bm{p}(n-1,t)
        + \begin{pmatrix}
            v_{1-} & 0 & 0 \\
            0 & v_{2-} & 0 \\
            0 & 0& v_{3-}
        \end{pmatrix}\bm{p}(n+1,t) 
         + \begin{pmatrix}
            d_1 & 0 & 0 \\
            u_{21} & d_2 & 0 \\
            u_{31} & 0 & d_3
        \end{pmatrix}\bm{p}(n,t),
    \end{align}
where $d_1 = -(u_{21} + u_{31} + v_{1+} + v_{1-})$, $d_2 = -(v_{2+} + v_{2-})$, $d_3 = -(v_{3+} + v_{3-})$, and $\bm{p}(n,t) = (p(n,1,t), p(n,2,t), p(n,3,t))^\top$.
Figure~\ref{fig: Transition diagrams}(b) represents the 2-random walk which has next-nearest-neighbor hoppings without internal degrees of freedom:
    \begin{equation}
        \frac{\mathrm{d}}{\mathrm{d}t} p(n,t)
        = a_2 p(n-2,t) + a_1 p(n-1,t) + b_1 p(n+1,t) + b_2 p(n+2,t) + d p(n,t)
    \end{equation}
where $d = -(a_1 + a_2 + b_1 + b_2)$.
Figure~\ref{fig: Transition diagrams}(c) represents the model which has nearest-neighbor hoppings with two internal degrees of freedom:
    \begin{align}
        \frac{\mathrm{d}}{\mathrm{d}t} \bm{p}(n,t)
        = \begin{pmatrix}
            0 & v_{+12} \\
            v_{+21} & 0
        \end{pmatrix} \bm{p}(n-1,t)
        + \begin{pmatrix}
            0 & v_{-12} \\
            v_{-21} & 0
        \end{pmatrix} \bm{p}(n+1,t) 
         + \begin{pmatrix}
            \tilde{d}_1 & u_{12} \\
            u_{21} & \tilde{d}_2
        \end{pmatrix} \bm{p}(n,t)
    \end{align}
where $\tilde{d}_1 = -(u_{21} + v_{+21} + v_{-21})$, $\tilde{d}_2 = -(u_{12} + v_{+12} + v_{-12})$, and $\bm{p}(n,t) = (p(n,1,t), p(n,2,t))^\top$.

\begin{figure}[t]
        \centering
        \includegraphics[width=172mm, bb=0 0 1550 1100, clip]{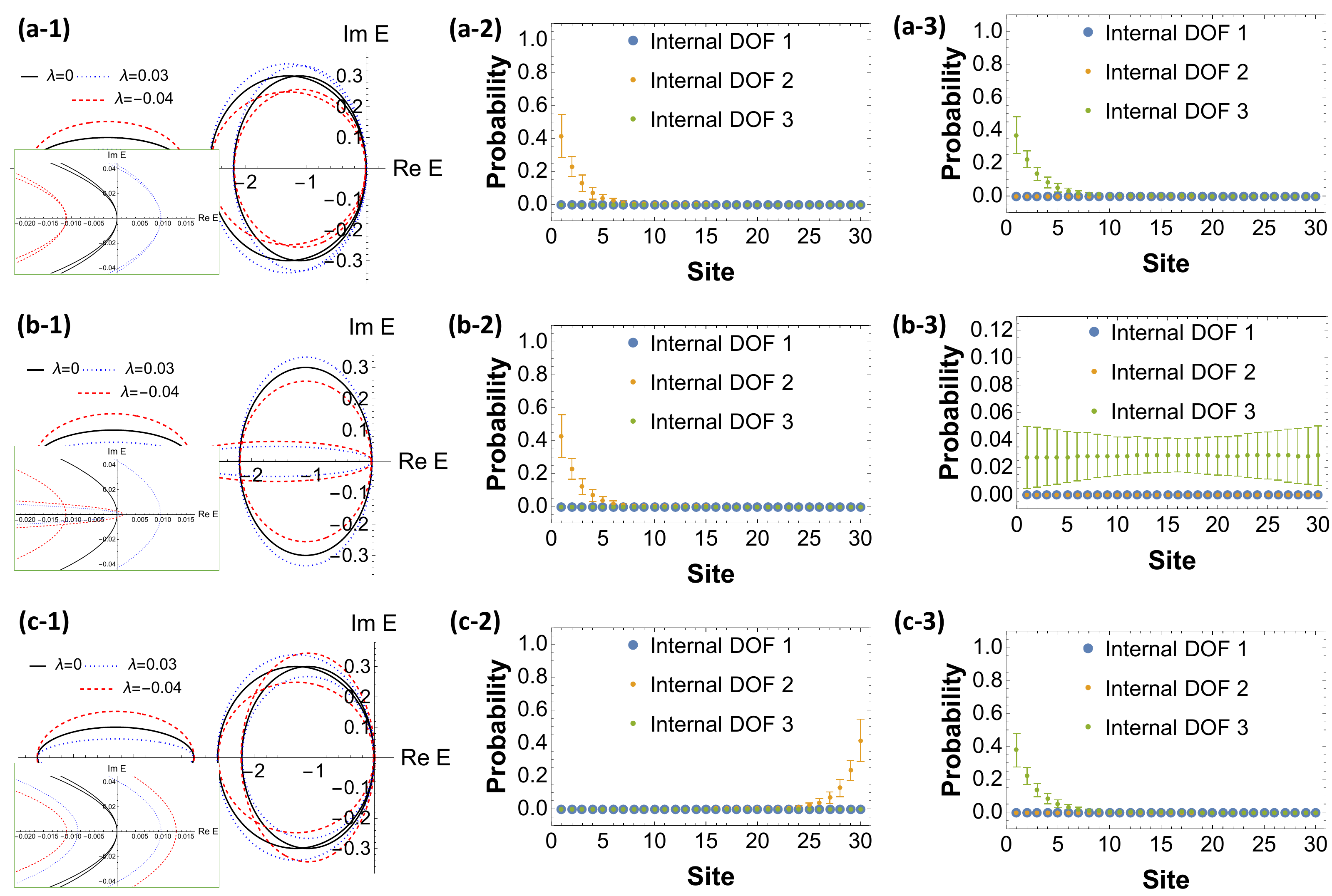}
        \caption{
        Numerical calculations of the winding number and the steady states under the OBC in the nonergodic model [Fig.~\ref{fig: Transition diagrams}(a)].
        The word DOF in legends is the abbreviation for degrees of freedom.
        (a-1), (b-1), (c-1) The PBC spectra of the original, and imaginary-gauge-transformed systems with $\lambda_+= 0.03$ and $\lambda_-= 0.04$.
        Black solid, red dashed, and blue dotted curves correspond to original, $\lambda_+$, and $\lambda_-$, respectively.
        The insets show the spectral curve around the zero spectrum.
        (a-2, 3), (b-2, 3), (c-2, 3) The OBC steady states with the system size $L = 20$.
        The data points (error bars) are the mean (variance) of the randomly generated realizations of off-diagonal disorder with the sample size $1000$.
        Parameters used are ($u_{21},u_{31},v_{1+},v_{1-},v_{2+},v_{2-},v_{3+},v_{3-}$) = (a) (1, 2, 0.7, 0.6, 0.4, 0.7, 0.5, 0.8), (b) (1, 2, 0.7, 0.6, 0.7, 0.4, 0.8, 0.8), (c) (1, 2, 0.7, 0.6, 0.7, 0.4, 0.5, 0.8).
        }
        \label{fig: Nonergodic windnum steadystate}
    \end{figure}

Firstly, we provide the winding number and the corresponding OBC steady state (Fig.~\ref{fig: Nonergodic windnum steadystate}).
The steady states of the nonergodic model are already shown in the main text.
As in the Lemma 2, the winding number is determined from the contribution of the sink components $G_2$ and $G_3$.
Moreover, as in the Lemma 3, the steady state has nonzero amplitude only in the single sink component $G_2$ or $G_3$.
We obtain the correspondence between the winding number and the number of the steady state by separately considering the steady state in each sink component.

\begin{figure}[t]
        \centering
        \includegraphics[width=140mm, bb=0 0 650 360, clip]{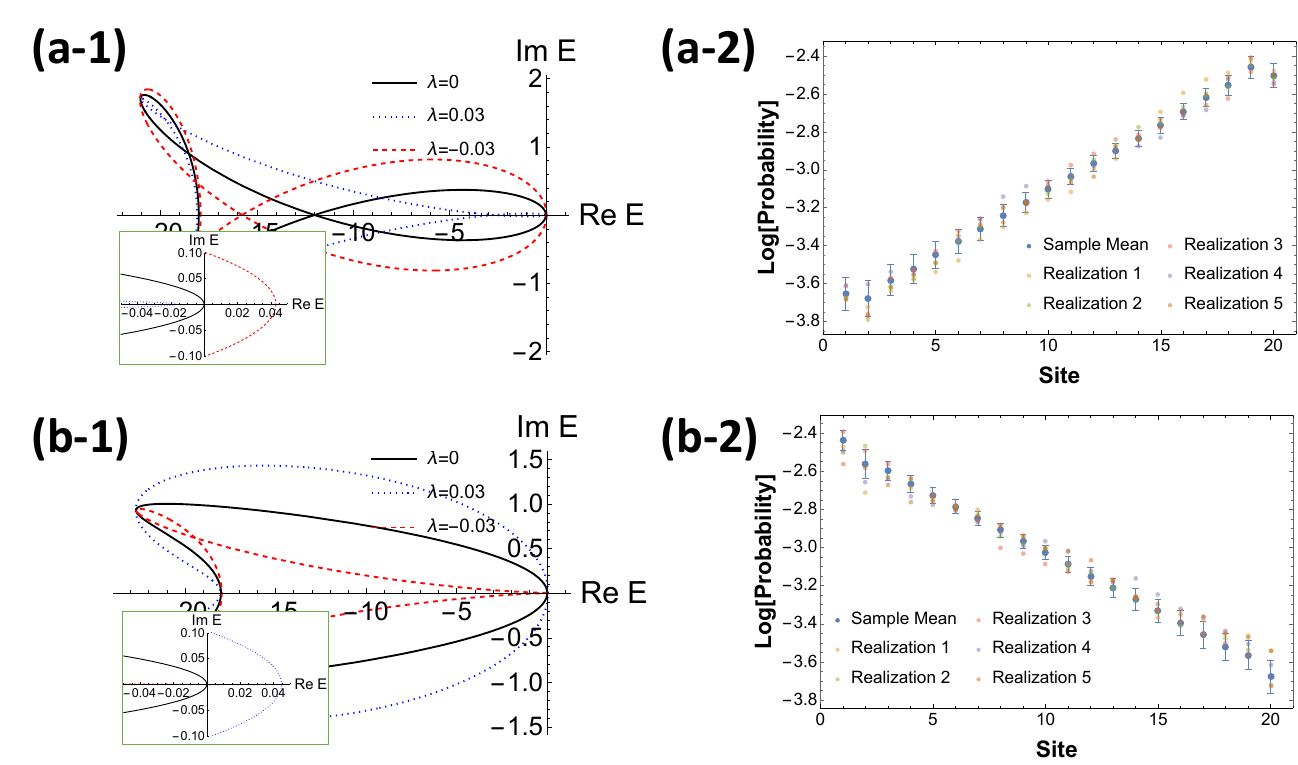}
        \caption{
        Numerical calculations of the winding number and the steady state under the OBC in the 2-random walk [Fig.~\ref{fig: Transition diagrams}(b)].
        (a-1), (b-1), (c-1) The PBC spectra of the original, and imaginary-gauge-transformed systems with $\lambda_\pm= 0.03$.
        Black solid, red dashed, and blue dotted curves correspond to original, $\lambda_+$, and $\lambda_-$, respectively.
        The insets show the spectral curve around the zero spectrum and we obtain the winding number as (a) $w = -1$, (b) $w = +1$.
        (a-2, 3), (b-2, 3), (c-2, 3) The OBC steady states with the system size $L = 20$.
        The data points (error bars) are the mean (variance) of the randomly generated realizations of off-diagonal disorder with the sample size $1000$.
        We also plot the five realizations of the steady state to confirm that the system has the robustness not only in the statistical meaning but also in the realization.
        The steady states are (mainly) localized to the (a) right or (b) left boundary corresponding to the winding number.
        Therefore, we confirm that the bulk-boundary correspondence also holds true in stochastic systems with longer-range hoppings.
        Parameters used are ($a_1, a_2, b_1, b_2$) = (a) (4,3,5,2), (b) (4,3,5,3).
        }
        \label{fig: 2RW windnum steadystate}
    \end{figure}

\begin{figure}[htbp]
        \centering
        \includegraphics[width=140mm, bb=0 0 650 360, clip]{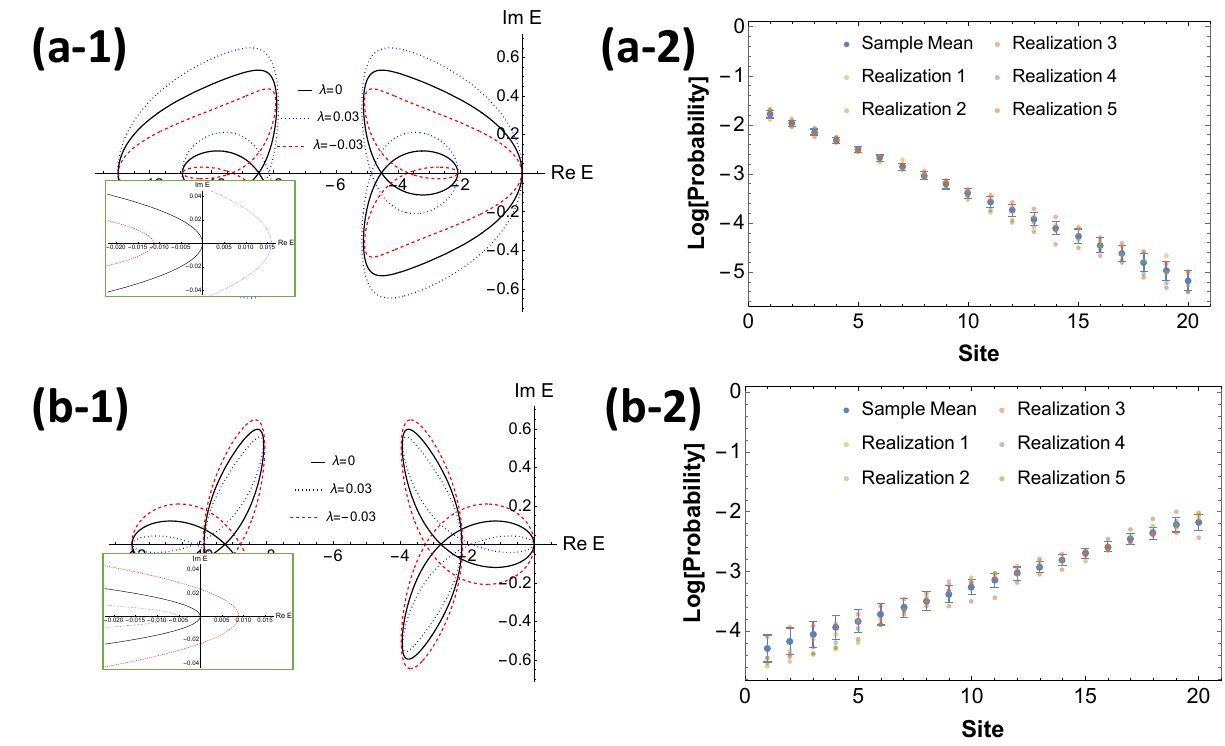}
        \caption{
        Numerical calculations of the winding number and the steady state under the OBC in the model with two internal degrees of freedom [Fig.~\ref{fig: Transition diagrams}(c)].
        (a-1), (b-1), (c-1) The PBC spectra of the original, and imaginary-gauge-transformed systems with $\lambda_\pm= 0.03$.
        Black solid, red dashed, and blue dotted curves correspond to original, $\lambda_+$, and $\lambda_-$, respectively.
        The insets show the spectral curve around the zero spectrum and we obtain the winding number as (a) $w = +1$, (b) $w = -1$.
        (a-2, 3), (b-2, 3), (c-2, 3) The OBC steady states with the system size $L = 20$.
        The data points (error bars) are the mean (variance) of the randomly generated realizations of off-diagonal disorder with the sample size $1000$.
        We also plot the five realizations of the steady state to confirm that the system has the robustness not only in the statistical meaning but also in the realization.
        We also plot the five realizations of the steady state to confirm the robustness of localization under the disorder.
        The steady states are localized to the (a) left or (b) right boundary corresponding to the winding number.
        Therefore, we confirm that the bulk-boundary correspondence also holds true with internal degrees of freedom.
        Parameters used are ($u_{21},u_{12},v_{+21},v_{-21},v_{+12},v_{-12}$) = (a) (0.7, 1.3, 2, 4, 3, 2), (b) (0.7, 1.3, 2, 4, 3, 1).
        }
        \label{fig: nn2state windnum steadystate}
    \end{figure}

\clearpage
Then, we see the results on the ergodic models to confirm that the bulk-boundary correspondence also holds true in the models with the longer hopping range (Fig.~\ref{fig: 2RW windnum steadystate}) and the internal degrees of freedom (Fig.~\ref{fig: nn2state windnum steadystate}).
As we can expect from the theorem, we confirm the correspondence between the winding number and the steady state, illustrating that our bulk-boundary correspondence is not affected by either the hopping range or the internal degrees of freedom.
Moreover, we also see that the exponential localization is not largely changed by the disorder since the logarithm of the spatial distribution is line-shaped and error bars are not so large.
We note that the steady states of the 2-random walk are localized at both ends (e.g. Fig.~\ref{fig: 2RW windnum steadystate} (a-2)) because of the next-nearest-neighbor hoppings, which seemingly does not represent the correspondence between the winding number and the number of the steady states.
However, we can recover the correspondence between the winding number and the number of the steady states since it is still possible to judge the direction of localization in the steady state based on the absolute value of the dominant non-Bloch wavenumber in the bulk.

\clearpage
\end{document}